\documentclass[a4paper,11pt]{article}
\pdfoutput=1 

\usepackage{jheppub} 
\usepackage[T1]{fontenc} 
\usepackage{amsmath}
\usepackage{slashed,graphicx,color,amsmath,amssymb}
\usepackage[mathscr]{eucal}
\usepackage{mathrsfs}
\usepackage{listings}
\usepackage{bm}
\usepackage[usenames,dvipsnames]{xcolor}

\title{\boldmath Diagonal Form Factors and Heavy-Heavy-Light Three-Point Functions at Weak Coupling}

\author[a]{Laszlo Hollo,}
\author[b]{ Yunfeng Jiang,}
\author[b]{ Andrei Petrovskii}

\affiliation[a]{MTA Lend\"ulet Holographic QFT Group, Wigner Research Centre for Physics H-1525 Budapest 114, P.O.B. 49, Hungary}
\affiliation[b]{Institut de Physique Th\'eorique,DSM, CEA, URA2306 CNRS\\Saclay, F-91191 Gif-sur-Yvette,France}

\emailAdd{hollo.laszlo@wigner.mta.hu}
\emailAdd{jinagyf2008@gmail.com}
\emailAdd{andreypetrovskij@gmail.com }

\newcommand{\tr}{\mathrm{Tr}\,}
\newcommand{\rT}{\mathrm{T}}
\newcommand{\rS}{\mathrm{S}}
\newcommand{\rH}{\mathrm{H}}
\newcommand{\rP}{\mathrm{P}}

\newcommand{\rK}{\mathrm{K}}
\newcommand{\rM}{\mathrm{M}}
\newcommand{\rN}{\mathrm{N}}
\newcommand{\rI}{\mathrm{I}}
\newcommand{\rE}{\mathrm{E}}
\newcommand{\rF}{\mathrm{F}}
\newcommand{\cF}{\mathcal{F}}
\newcommand{\bOmega}[1]{\textcolor{blue}{\Omega_{#1}}}
\newcommand{\Tr}{\mathrm{Tr}\,}

\newcommand{\beq}{\begin{equation}}
\newcommand{\eeq}{\end{equation}}
\newcommand{\ur}[1]{(\ref{#1})}

\abstract{In this paper we consider a special kind of three-point functions of HHL type at weak coupling in ${\cal N}=4$ SYM theory and analyze its volume dependence. At strong coupling this kind of three-point functions were studied recently by Bajnok, Janik and Wereszczynski \cite{Bajnok:2014sza}. The authors considered some cases of HHL correlator in the $\mathfrak{su}(2)$ sector and, relying on their explicit results, formulated a conjecture about the form of the volume dependence of the symmetric HHL structure constant to be valid at any coupling up to wrapping corrections. In order to test this hypothesis we considered the HHL correlator in $\mathfrak{su}(2)$ sector at weak coupling and directly showed that, up to one loop, the finite volume dependence has exactly the form proposed in \cite{Bajnok:2014sza}. Another side of the conjecture suggests that computation of the symmetric structure constant is equivalent to computing the corresponding set of infinite volume form factors, which can be extracted as the coefficients of finite volume expansion. In this sense, extracting appropriate coefficients from our result gives a prediction for the corresponding infinite volume form factors.}

\begin{document}
\maketitle
\flushbottom

\section{Introduction}

It's strongly believed that ${\cal N}=4$ SYM theory is integrable. There are a huge amount of supportive evidences in the literature starting from \cite{MinahanZarembo}, where the relation between spin chain Hamiltonian and the dilatation operator of the theory was established, for detailed review, see \cite{BigReview}. Among one of the most exciting achievements revealing the power of the integrability is the so-called Quantum Spectral Curve (QSC) approach \cite{QSC,QSC2} which provides efficient technique for computing anomalous dimension of a gauge invariant operators of the theory.

Due to conformal symmetry, the computation of general correlation functions of ${\cal N}=4$ SYM can be reduced to computation of two- and three-point functions by means of the operator product expansion (OPE). In this sense these two quantities are the fundamental blocks of the theory. Two-point functions of gauge invariant operators, again due to the conformal symmetry, are completely defined by the value of the corresponding anomalous dimension, and thus can be computed in terms of QSC technique. However for the moment there is no such an efficient analogue for computing three-point functions. Nevertheless significant progress have been done in this direction starting from \cite{Okuyama-Tseng,Alday:3pt} and later developed systematically in \cite{EGSV:Tailoring1}, where the three-point functions of the $\mathfrak{su}(2)$ sector at tree level were considered and, using the mapping between spin chains and gauge invariant operators, the computation of the structure constants was reduced to computation of the scalar products between two off-shell Bethe-states. This method was improved in \cite{Foda:3ptdeterminant} for some special configuration, which allows to express the structure constant in terms of on-shell/off-shell scalar products and can be computed by Slavnov determinants. Later this result was extended to one loop \cite{GV,Jiang:OneLoop} and other rank-one sectors also were investigated \cite{Pedro:sl2,Caetano:fermion}. The great advancement was made very recently in \cite{Basso:3ptf}, where the authors proposed all-loop procedure for the structure constant in ${\cal N}=4$ SYM. Another interesting direction inspired by the light-cone string field theory was initiated in \cite{NeumCoef:Zoli}, where the authors interpreted the OPE coefficients as ``generalized Neumann coefficients'' and proposed a set of bootstrap axioms for these generalized Neumann coefficients. In parallel, the spin vertex approach was developed in \cite{spvertex3,spvertex,spvertex2} which can be seen as a weak coupling counterpart of the string vertex.

Apart from direct approach, there is another method for the computation of three-point functions by means of their relations with the form factors. This approach was initiated in \cite{Klose:FF,KloseMc2}, where the set of the axioms for the world-sheet form factor of the light-cone gauge fixed $AdS_5\times S^5$ string theory was proposed.
Recently in \cite{Bajnok:2014sza} the symmetric HHL
(heavy-heavy-light) correlator at strong coupling of the $\mathfrak{su}(2)$ sector was considered. The proposals for computing the HHL three-point functions were first
formulated in \cite{Zarembo:2010rr,Costa:2010rz}. The authors of \cite{Bajnok:2014sza} showed that the prescription in \cite{Zarembo:2010rr} was inadequate and proposed an improved prescription. Using this new prescription the authors computed several examples of the three-point correlators and showed that their volume dependence
exactly coincide with the finite volume structure of the appropriate form factor. Guided by this result they proposed a conjecture that this finite volume dependence should hold at any coupling of the theory.

The conjecture consists two parts. First, the finite volume dependence (neglecting wrapping) of a symmetric HHL structure constant is completely encoded into diagonal minors of the Gaudin determinant of a heavy state and have the following form dictated by results of form factor theory (\cite{Pozsgay:2007gx}):
\beq
\begin{split}
&C_{HHL}=\frac{f^{{\cal O}} + \sum\limits_i \rho_N(\{i\}) f^{\cal O}(1,...,\hat{i},...,N)+ \sum\limits_{i,j}\rho_N(\{i,j\})f^{\cal O}(1,...,\hat{i},...,\hat{j},...,N)+...}{\rho_N(\{1,...,N\})},
\end{split}
\label{fvf}
\eeq
and $\rho_N(\{i,j,...\})$'s are diagonal minors of the determinant:
\beq
\rho_N(\{1,...,N\})=\det\limits_{j,k}\Bigg(\frac{\partial}{\partial u_j}\Big(L p_k(u) + \sum\limits_{l\neq k} \frac{1}{i}\log S(u_l,u_k)\Big)\Bigg).
\eeq
Second part of the conjecture, also coming from the form factor approach, is that the coefficients $f_{k}^{\cal O}$ in the expansion \ur{fvf} are diagonal infinite volume form factors of a light operator ${\cal O}$. In this paper we test the first part of the conjecture at weak coupling limit and show that the form \ur{fvf} of the finite volume dependence holds at tree level as well as at one loop. At the same time the second part of the conjecture allowed us to predict expressions for diagonal infinite volume form factors by extracting corresponding coefficients from the expansion \ur{fvf}.

The structure of the paper is the following. In section 2 we set up the problem. In section 3, we will describe in detail the map between the field theory operator $\mathcal{O}_\alpha$ and the spin chain operator $\hat{O}_\alpha$. In section 4, we review the integrability tools that we need, which includes algebraic Bethe ansatz (ABA), the scalar products between Bethe states and the solution of quantum inverse scattering problem (QISP). In section 5, we review the form factor formalism and the finite volume corrections. In section 6 and 7, we give a proof of the finite volume structure of the three-point function conjectured in \cite{Bajnok:2014sza} at tree level. In section 8 we compute some examples of the symmetric HHL correlator and list expressions for the corresponding infinite volume form factors. In section 9 we give general arguments proving that the same finite volume structure holds as well at one loop. Finally we conclude and discuss future directions to pursue in section 10.

\section{Set-up}

The central object of our paper is the symmetric HHL three-point correlation function of gauge invariant operators,
by which we mean that two heavy\footnote{Although we call it "heavy" the results obtained in this paper are valid
for small values of $L$ as well} (the length $L \gg 1$) operators are conjugated to each other and the third operator has a few number of sites. The technique, which we are going to use, is due to the relation between
the gauge invariant operators and the spin chains.
Following \cite{Roiban:3pt}, the symmetric structure constant can be mapped to diagonal matrix element of a spin chain operator.
We take the heavy operators to be the eigenstates of the
$\mathfrak{su}(2)$ sector dilatation operator and conjugated to each other. They are constructed from the following scalar fields

\begin{align}
\mathcal{O}_{1}:\{Z,X\},\quad\qquad\mathcal{O}_{\bar{1}}:\{\bar{Z},\bar{X}\}.
\end{align}
At tree-level, we consider operators with definite one-loop anomalous
dimension \cite{EGSV:Tailoring1}. The one-loop dilatation operator in the $\mathfrak{su}(2)$
sector is the Heisenberg XXX$_{1/2}$ spin chain Hamiltonian whose
eigenstates can be constructed by Bethe Ansatz techniques. Since we
are considering diagonal matrix elements, the wave functions of the
two operators are conjugate to each other. The third operator will be denoted
hereafter by $\mathcal{O}_{\alpha}$. In this paper, we consider $\mathcal{O}_\alpha$ in the compact sector. Only the zero $R$-charge terms give
non-zero contribution.
Therefore, we are interested in the operators of the following form
\begin{align}
\mathrm{Tr}\, Z\bar{Z},\quad\mathrm{Tr}\, X\bar{X},\quad\mathrm{Tr}\, XZ\bar{Z}\bar{X},\quad\cdots
\end{align}
This kind of operators are in the $\mathfrak{so}(4)$ sector of $\mathcal{N}=4$
SYM theory. In addition, we require $\mathcal{O}_{\alpha}$ have definite
anomalous dimension. The three-point function is fixed by conformal
symmetry up to the structure constant $C_{\alpha}$
\begin{align}
\langle\mathcal{O}_{1}(x_{1})\mathcal{O}_{\alpha}(x_{2})\mathcal{O}_{\bar{1}}(x_{3})\rangle=\frac{L^{2}L_{\alpha}}{N_{c}}\frac{\mathcal{N}_{1}\sqrt{\mathcal{N}_{\alpha}}C_{\alpha}}{|x_{12}|^{\Delta_{12}}|x_{13}|^{\Delta_{13}}|x_{23}|^{\Delta_{23}}}
\end{align}
where
\begin{align}
x_{ij}^{\mu}=x_{i}^{\mu}-x_{j}^{\mu},\qquad\Delta_{ij}=\frac{1}{2}(\Delta_{i}+\Delta_{j}-\Delta_{k}),
\end{align}
and $L$ is the length of $\mathcal{O}_{1}$ while $L_{\alpha}$ is
the length of the operator $\mathcal{O}_{\alpha}$. The two-point
functions are normalized as
\begin{align}
\langle\mathcal{O}_{1}(x_{1})\mathcal{O}_{\bar{1}}(x_{2})\rangle=\frac{L\mathcal{N}_{1}}{|x_{12}|^{2\Delta_{1}}},\quad\langle\mathcal{O}_{\alpha}(x_{1})\overline{\mathcal{O}}_{\alpha}(x_{2})\rangle=\frac{L_{\alpha}\mathcal{N}_{\alpha}}{|x_{12}|^{2\Delta_{\alpha}}}
\end{align}
The structure constant $C_{\alpha}$ can be expressed in terms of
correlation functions of the Heisenberg spin chain
\begin{align}
C_{\alpha}=\frac{\langle\mathbf{u}|\hat{O}_{\alpha}|\mathbf{u}\rangle}{\langle\mathbf{u}|\mathbf{u}\rangle}\label{C}
\end{align}
where $|\mathbf{u}\rangle\equiv|\{u_{1},\cdots,u_{N}\}\rangle$ denotes
an on-shell Bethe state, corresponding to the heavy operator, and $\hat{O}_{\alpha}=\hat{O}_{\alpha}(\sigma_{i}^{\pm},\sigma_{i}^{z})$
is an operator made of local spin operators. In this way, the computation
of three-point function in planar $\mathcal{N}=4$ SYM theory is recast
into the calculation of correlation functions in the Heisenberg spin
chain.

\section{From field theory correlation functions to spin chain matrix elements}

\label{eq:CFtoFF} 
\begin{figure}[h!]
\begin{centering}
\includegraphics[scale=0.4]{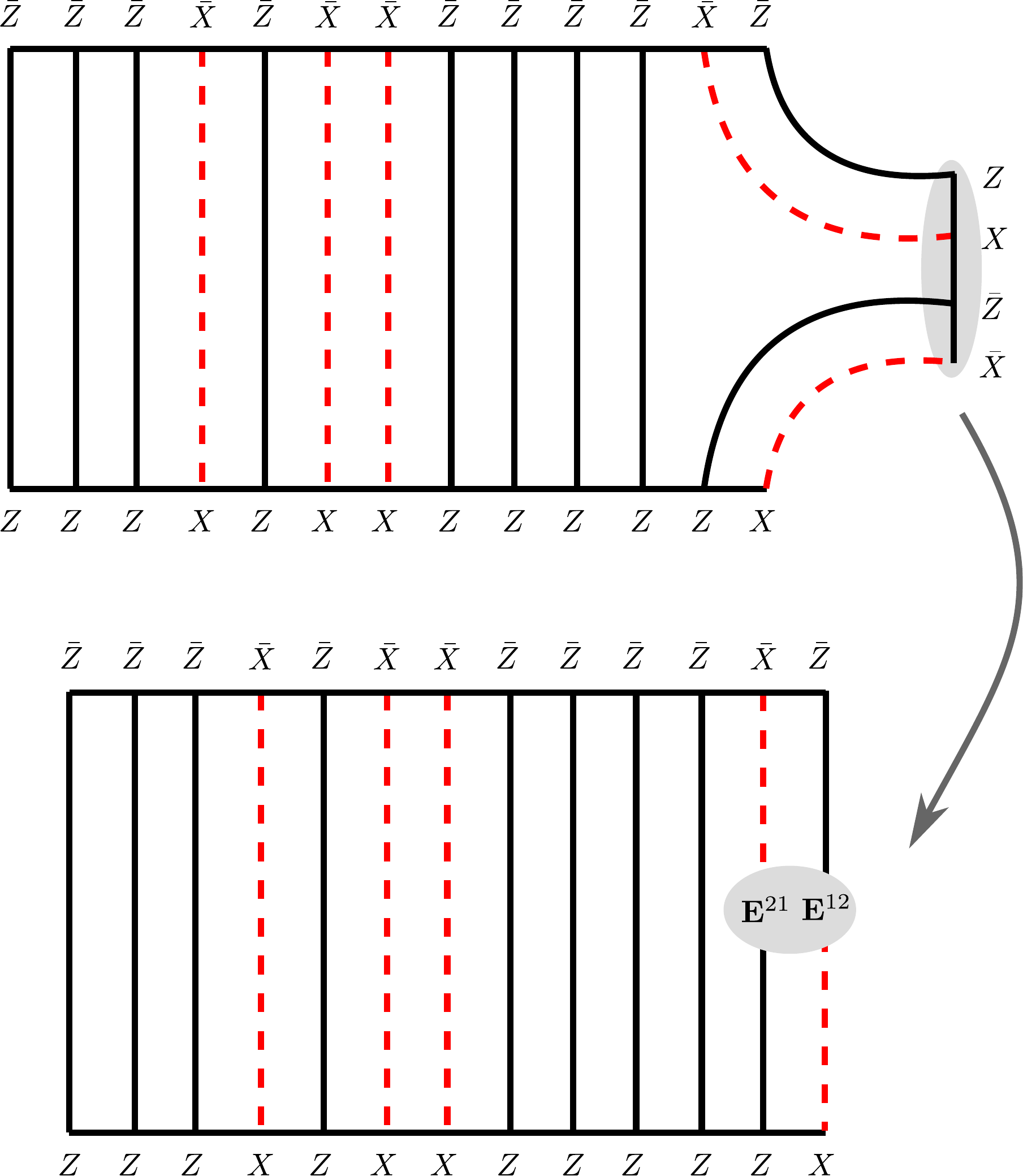} \caption{An example of the mapping between field operators to spin chain operators.
The operator in the field theory $\mathcal{O}=\tr ZX\bar{Z}\bar{X}$
is mapped to the spin chain operator $\hat{O}=\rE_{n}^{21}\rE_{n+1}^{12}$.}

\par\end{centering}

\centering{}\label{FF}
\end{figure}

In this section, we summarize how to write the
field theoretic operators $\mathcal{O}_{\alpha}$ in terms of spin
chain operators $\hat{O}_{\alpha}$. Let us introduce the following
notation
\begin{align}
Z\equiv\phi^{0},\quad X\equiv\phi^{1},\quad\bar{Z}\equiv\bar{\phi}^{0},\quad\bar{X}\equiv\bar{\phi}^{1}.
\end{align}
The light operator is the linear combination of the single trace operators
$\tr\phi^{i_{1}}\bar{\phi}^{i_{2}}\bar{\phi}^{i_{3}}\cdots$. By planarity,
only operators of the following form will contribute to the three-point
function
\begin{align}
\mathrm{Tr}\,\phi_{1}^{i_{1}}\cdots\phi_{l}^{i_{l}}\bar{\phi}_{l+1}^{j_{l}}\cdots\bar{\phi}_{2l}^{j_{1}},\quad i_{k},j_{k}=0,1.\label{phi}
\end{align}
where the indices $1,\cdots,2l$ denotes the position on the third
spin chain and $2l=L_{\alpha}$ is the length of the third operator.
The zero $R$-charge condition is given by
\begin{align}
\sum_{n=1}^{l}(i_{n}-j_{n})=0.
\end{align}
It is not hard to see that the operator (\ref{phi}) can be mapped
to the following spin operator \cite{Roiban:3pt}
\begin{align}
\mathrm{Tr}\,\phi_{1}^{i_{1}}\cdots\phi_{l}^{i_{l}}\bar{\phi}_{l+1}^{j_{l}}\cdots\bar{\phi}_{2l}^{j_{1}}\longrightarrow\rE_{n+l-1}^{i_{1}+1,j_{1}+1}\cdots\rE_{n}^{i_{l}+1,j_{l}+1}.\label{mapping}
\end{align}
where the indices $n,\cdots,n-l+1$ denote the positions on the long
spin chain, and the operators $\rE_{n}^{ab}$ are the basis $2\times2$
matrices $(\rE_{n}^{ab})_{ij}=\delta_{ai}\delta_{bj}$ in the local
quantum space $\mathcal{H}_{n}=\mathbb{C}^{2}$. The operators $\rE_{n}^{ab}$
are related to the local spin operators as follows
\begin{align}
\rE_{n}^{11}\equiv\frac{1}{2}(\mathbb{I}+\sigma_{n}^{z}),\quad\rE_{n}^{12}\equiv\sigma_{n}^{+},\quad\rE_{n}^{21}\equiv\sigma_{n}^{-},\quad\rE_{n}^{22}\equiv\frac{1}{2}(\mathbb{I}-\sigma_{n}^{z}).\label{Eab}
\end{align}
Here $\sigma_{n}^{\pm},\,\sigma_{n}^{z}$ are the usual Pauli matrices acting on the space spanned by $|\!\uparrow\rangle$ and $|\!\downarrow\rangle$.
By the mapping (\ref{mapping}), we can translate the field theory
operators into the spin operators. As an example, we consider the
Konishi operator
\begin{align}
\mathcal{O}_{K}=\mathrm{Tr}X\bar{X}+\mathrm{Tr}Y\bar{Y}+\mathrm{Tr}Z\bar{Z}.
\end{align}
Since the heavy operators are in the $\mathfrak{su}(2)$ sector, the
contraction with $Y$ and $\bar{Y}$ are zero and can be neglected.
The Konishi operator can be mapped to the following spin operator
\begin{align}
\mathcal{O}_{K}\longrightarrow\hat{O}_{K}=\rE_{n}^{11}+\rE_{n}^{22}=\mathbb{I}.
\end{align}
Therefore we see that at tree level the structure constant, with the
light operator being the Konishi operator, is trivial
\begin{align}
C_{K}=\frac{\langle\mathbf{u}|\hat{O}_{K}|\mathbf{u}\rangle}{\langle\mathbf{u}|\mathbf{u}\rangle}=1.
\end{align}
In order to obtain non-trivial structure constant, the light operator
must have at least $L_{\alpha}=4$. This corresponds to the insertion
of two spin operators between the Bethe states. An example for length-4
operator is given in Fig.\,\ref{FF}.

The main focus of the current paper is the length-4 field theory operators,
which corresponds to operators in the spin chain which acts on two
neighboring sites. These are the simplest non-trivial cases which
can be studied thoroughly. We use $L_{\alpha}$ and $l_{s}$ to denote
the lengths of field theory operators and spin chain operators, it
is obvious that $L_{\alpha}=2l_{s}$.

In order to have non-zero diagonal matrix element in (\ref{C}), the
spin chain operator should not alter the total spin of the state it acts on. At $l_{s}=1$, there are two independent operator,
satisfying this condition, the identity and
\begin{align}
o_{1}(n)=\rE_{n}^{11}.\label{eq:o1}
\end{align}
For $l_{s}=2$, there are six independent operators: the unity, $o_{1}(n)$, $o_{1}(n+1)$
\footnote{Due to translation invariance of the spin chain at the level of correlation functions these two operators obviously coincide.} and the other three operators are the following
\begin{align}
o_{2}^{1}(n)=\rE_{n}^{11}\rE_{n+1}^{11},\quad o_{2}^{2}(n)=\rE_{n}^{12}\rE_{n+1}^{21}.\quad o_{2}^{3}(n)=\rE_{n}^{21}\rE_{n+1}^{12}.\label{eq:o2}
\end{align}
For later convenience we also introduce the following operator
\begin{align}
o_2^4=\rE_{n}^{22}\rE_{n+1}^{22}=(\mathbb{I}-\rE_{n}^{11})(\mathbb{I}-\rE_{n+1}^{11})=\mathbb{I}-o_{1}(n)-o_{1}(n+1)+o_{2}^{1}(n).\label{eq:o4_exp}
\end{align}
In what follows, we will study in detail the diagonal matrix elements
of the spin operators $o_{1}$ and $o_{2}^{i}$ $(i=1,2,3,4)$.

\section{The Algebraic Bethe Ansatz approach}

\label{sec:ABA} In this section, we review the algebraic Bethe Ansatz
(ABA) approach to the correlation functions in the XXX$_{1/2}$ Heisenberg
spin chain, see \cite{Kitanie:FF,Maillet:QISP,Kitanine:Review} and
references therein. This approach is based on two main elements: the
Slavnov determinant formula and the solution of the Quantum Inverse
Scattering Problem (QISP). The Slavnov formula states that the scalar
product of any Bethe state with an on-shell one can be written in
terms of a determinant. The solution of the QISP enable us to express
the local spin operators in terms of the elements of the monodromy
matrix, which are non-local operators acting on the spin chain.

\subsection{Algebraic Bethe Ansatz}

In this subsection we give a brief review of the algebraic Bethe Ansatz
method which also serves to fix our conventions. For more detailed
and pedagogical discussions we refer the readers to \cite{Faddeev:ABA}.
The central object in ABA is the quantum $R$-matrix. For the Heisenberg
XXX$_{1/2}$ spin chain, the $R$-matrix is given by
\begin{align}
R_{ab}(u)=u\,\mathrm{I}_{ab}+i\,\rP_{ab}=\left(\begin{array}{cccc}
u+i & 0 & 0 & 0\\
0 & u & i & 0\\
0 & i & u & 0\\
0 & 0 & 0 & u+i
\end{array}\right)_{ab}
\end{align}
where $\mathrm{I}_{ab}$ and $\rP_{ab}$ are identity and permutation
operators acting on the space $\mathbb{C}_{a}^{2}\otimes\mathbb{C}_{b}^{2}$,
respectively. It has a complex parameter which is usually referred
to as the spectral parameter. Consider a spin chain of length $L$.
At each site of the spin chain we define the Lax matrix
\begin{align}
L_{an}(u-\theta_{n})=R_{an}(u-\theta_{n}-i/2),\quad n=1,\cdots,L
\end{align}
where $n$ and $a$ denote the quantum space and auxiliary space,
respectively. The parameter $\theta_{n}$ is called the inhomogeneity
associated to the site $n$. In the case of the homogeneous Heisenberg
spin chain all the inhomogeneities are set to be zero, $\theta_{n}=0$.
The Lax operator obeys an important property that it satisfies the
following braiding relation, called the $RLL$ relation
\begin{align}
R_{ab}(u-v)L_{an}(u)L_{bn}(v)=L_{bn}(v)L_{an}(u)R_{ab}(u-v).\label{eq:RLL}
\end{align}
Taking the ordered product of Lax operators at all sites, we obtain
the monodromy matrix
\begin{align}
T_{a}(u)=\prod_{n=1}^{L}L_{an}(u-\theta_{n}).\label{eq:mono}
\end{align}
It can be represented in the auxiliary space as a $2\times2$ matrix
\begin{align}
T(u)=\left(\begin{array}{cc}
A(u) & B(u)\\
C(u) & D(u)
\end{array}\right).
\end{align}
By construction, the monodromy matrix is a non-local object and its
elements are operators which act on the whole spin chain. From the
$RLL$ relation (\ref{eq:RLL}) and the definition of the monodromy
matrix (\ref{eq:mono}), one can show that the monodromy matrix satisfies
a similar relation, called the $RTT$ relation
\begin{align}
R_{ab}(u-v)T_{a}(u)T_{b}(v)=T_{b}(v)T_{a}(u)R_{ab}(u-v).
\end{align}
Once written in terms of the components, the $RTT$ relation leads
to a quadratic algebra,
between the ABA operators $A$, $B$, $C$, $D$. The most relevant
relations for us are
\begin{align}
A(v)B(u)= & \, f(u-v)B(u)A(v)+g(v-u)B(v)A(u),\label{eq:Yangian}\\
D(v)B(u)= & \, f(v-u)B(u)D(v)+g(u-v)B(v)D(u),\nonumber \\
{}[C(v),B(u)]= & \, g(u-v)[A(v)D(u)-A(u)D(v)]\nonumber
\end{align}
where the functions $f(u)$ and $g(u)$ are
\begin{align}
f(u)=\frac{u+i}{u},\qquad g(u)=\frac{i}{u}.\label{eq:fg}
\end{align}

All the conserved charges of the system are encoded into the transfer
matrix, which is defined to be the trace of monodromy matrix in the
auxiliary space
\begin{align}
\mathcal{T}(u)=\mathrm{Tr}_{a}\, T(u)=A(u)+D(u).
\end{align}
The eigenstates of the transfer matrix simultaneously diagonalize
all the conserved charges, including the Hamiltonian. In order to
diagonalize it, one starts with a ferromagnetic reference state, called
the psuedovacuum $|\Omega\rangle=|\uparrow^{L}\rangle$, such that
\begin{align}
A(u)|\Omega\rangle=a(u)|\Omega\rangle,\quad D(u)|\Omega\rangle=d(u)|\Omega\rangle,\quad C(u)|\Omega\rangle=0.
\end{align}
For later convenience, we define the Baxter polynomials
\begin{align}
Q_{\bm{\theta}}(u)=\prod_{n=1}^{L}(u-\theta_{n}),\qquad Q_{\bm{\theta}}^{\pm}=Q_{\bm{\theta}}(u\pm i/2).
\end{align}
Then we have
\begin{align}
a(u)=Q_{\bm{\theta}}^{+}(u),\qquad d(u)=Q_{\bm{\theta}}^{-}(u).
\end{align}
The eigenstates of the transfer matrix are generated by the action
of a sequence of $B$-operators on the pseudovacuum
\begin{align}
|\mathbf{u}\rangle=B(u_{1})\cdots B(u_{N})|\Omega\rangle
\end{align}
provided that the rapidities $\mathbf{u}=\{u_{1},\cdots,u_{N}\}$
satisfy the Bethe Ansatz equation (BAE)
\begin{align}
\frac{a(u_{k})}{d(u_{k})}=\prod_{{j=1\atop j\ne k}}^{N}\frac{u_{k}-u_{j}+i}{u_{k}-u_{j}-i},\qquad k=1,\cdots,N.\label{eq:BAE_spinchain}
\end{align}
These elementary excitations are commonly referred to as magnons.
If the rapidities satisfy (\ref{eq:BAE_spinchain}), the corresponding
Bethe state $|\mathbf{u}\rangle$ is called \emph{on-shell}, otherwise
it is called \emph{off-shell}. The on-shell Bethe states diagonalize
the transfer matrix
\begin{align}
\mathcal{T}(u)|\mathbf{u}\rangle=t_{\mathbf{u}}(u)|\mathbf{u}\rangle
\end{align}
with the eigenvalue
\begin{align}
t_{\mathbf{u}}(u)=a(u)\frac{Q_{\mathbf{u}}(u-i)}{Q_{\mathbf{u}}(u)}+d(u)\frac{Q_{\mathbf{u}}(u+i)}{Q_{\mathbf{u}}(u)}.\label{trans}
\end{align}
Here, similarly we have defined the Baxter polynomial for the rapidities
\begin{align}
Q_{\mathbf{u}}(u)=\prod_{k=1}^{N}(u-u_{k}).
\end{align}
In this paper, we will compute the matrix elements of spin chain operators
between two Bethe states. To this end, it is important to have a manageable
expression for the scalar products between Bethe states.

\subsection{Slavnov determinant and Gaudin norm}

The scalar product of Bethe states is defined as
\begin{align}
\langle\mathbf{v}|\mathbf{u}\rangle=\langle\Omega|\prod_{j=1}^{N}C(v_{j})\prod_{k=1}^{N}B(u_{k})|\Omega\rangle.\label{SP}
\end{align}
The Slavnov theorem \cite{Slavnov:determinant} states that, if one
of the Bethe states, say $|\mathbf{u}\rangle$ is on-shell, the scalar
product (\ref{SP}) is given by
\begin{align}
\langle\mathbf{v}|\mathbf{u}\rangle=\prod_{j=1}^{N}a(v_{j})d(u_{j})\mathscr{S}_{\mathbf{u},\mathbf{v}},
\end{align}
where
\begin{align}
\mathscr{S}_{\mathbf{u},\mathbf{v}}=\frac{\det_{jk}\Omega(u_{j},v_{k})}{\det_{jk}\frac{1}{u_{j}-v_{k}+i}}\label{sp}
\end{align}
The matrix element $\Omega(u_{j},v_{k})$ is given by
\begin{align}
\Omega(u_{j},v_{k})= & \, t(u_{j}-v_{k})-e^{2ip_{\mathbf{u}}(v_{k})}t(v_{k}-u_{j})\label{Slavnov}\\
= & \, i\frac{(u_{j}-v_{k}-i)-(u_{j}-v_{k}+i)e^{2ip_{\mathbf{u}}(v_{k})}}{(u_{j}-v_{k})[(u_{j}-v_{k})^{2}+1]}\nonumber
\end{align}
where $p_{\mathbf{u}}(v)$ is the pseudomomentum
\begin{align}
e^{2ip_{\mathbf{u}}(u)}\equiv\frac{d(u)}{a(u)}\frac{Q_{\mathbf{u}}(u+i)}{Q_{\mathbf{u}}(u-i)}.
\end{align}
and $t(u)$ is
\begin{align}
t(u)=\frac{1}{u}-\frac{1}{u+i}.
\end{align}
In terms of the pseudomomentum, the Bethe Ansatz equation takes a
particularly simple form
\begin{align}
e^{2ip_{\mathbf{u}}(u_{k})}=-1.
\end{align}

In the Slavnov formula, the rapidities $\{\mathbf{v}\}$ can be any
set of complex numbers. Let us now consider the case when some of
the $v_{k}$'s coincide with the Bethe roots, namely $v_{k}=u_{k}$
for some $k$. The matrix element reads as
\begin{align}
\Omega(u_{j},u_{k})=\frac{2i}{(u_{j}-u_{k})^{2}+1}-i\delta_{jk}\left(\sum_{l=1}^{N}\frac{2}{(u_{j}-u_{l})^{2}+1}-\sum_{m=1}^{L}\frac{1}{(u_{j}-\theta_{m})^{2}+1/4}\right)
\end{align}

When all $\{\mathbf{v}\}$ coincide with $\{\mathbf{u}\}$, the scalar
product corresponds to the norm of the Bethe state. This norm is given
by the Gaudin formula which takes the form of a Jacobian determinant.
Let us define the following quantity
\begin{align}
\Phi_{k}(\mathbf{u})=p(u_{k})L-i\sum_{j\ne k}^{N}\log S(u_{k}-u_{j})\label{eq:Phik}
\end{align}
where the momentum and the two-body scattering matrix of the Heisenberg
spin chain are
\begin{align}
p(u)=-\frac{i}{L}\log\frac{a(u)}{d(u)},\qquad S(u-v)=\frac{u-v-i}{u-v+i}.\label{eq:pu}
\end{align}
In terms of the function $\Phi_{k}(\mathbf{u})$, the Bethe Ansatz
equation (\ref{eq:BAE_spinchain}) reads as
\begin{align}
\Phi_{k}(\mathbf{u})=2\pi I_{k},\qquad k=1,...,N\label{eq:BAE_Phik}
\end{align}
where $I_{k}\in\mathbb{Z}$ are the Bethe quantum numbers. The Jacobian
for the change of variables between $u_{k}$ and $I_{k}$ is given
by
\begin{align}
\rho_{N}(\{1,...,N\})=\left|\frac{\partial\Phi_{k}(\mathbf{u})}{\partial u_{j}}\right|=\det_{jk}i\,\Omega(u_{j},u_{k}).\label{eq:rhon}
\end{align}
The norm of an on-shell $N$-magnon Bethe states is proportional to
$\rho_{N}(\{1,\cdots,N\})$, explicitly
\begin{align}
\langle\mathbf{u}|\mathbf{u}\rangle=\left(\prod_{j=1}^{N}a(u_{j})d(u_{j})\prod_{j<k}^{N}\frac{1+(u_{j}-u_{k})^{2}}{(u_{j}-u_{k})^{2}}\right)\,\rho_{N}(\{1,...,N\})\label{eq:norm}
\end{align}
We will see below that the Jacobian $\rho_{N}(\{1,\cdots,N\})$ and
its sub-diagonal counterparts take into account all the finite volume
corrections of the diagonal matrix elements.

\subsection{The solution of quantum inverse scattering problem}

In the ABA approach, the idea to compute the diagonal matrix element $\langle\mathbf{u}|\hat{O}_{n}|\mathbf{u}\rangle$
is to act the spin chain operator $\hat{O}_{n}$ on the bra state,
so that the ket Bethe state $|\mathbf{u}\rangle$ is left on-shell.
The operator $\hat{O}_{n}$ is typically a multilocal operator in
the sense that it acts on a finite interval of the spin chain. In
order to apply the Slavnov formula, the bra state also needs to be
a Bethe state, although we do not require it to be on-shell. The solution
of the quantum inverse scattering problem (QISP) relates the local
spin operators to the matrix elements of the monodromy matrix. These
latter are non-local operators. In addition, from the Yangian algebra
(\ref{eq:Yangian}), it is clear that the action of $A$, $B$, $C$,
$D$ operators on a Bethe state always gives a sum over Bethe states.
Therefore, using the solution of QISP we can translate any local operator
$\hat{O}_{n}$ into a sequence of ABA operators, and the state $\langle\mathbf{u}|\hat{O}_{n}$
can be written as a sum of dual Bethe states. This enable us to apply
the Slavnov formula discussed in the last subsection. We present the
main statement of the solution of QISP in this subsection and refer
to \cite{Maillet:QISP} for the proof and details.

Let $\rE_{n}^{ab}$ ($a,b=1,2$) be the operators defined in (\ref{Eab})
which act on the local quantum space $\mathcal{H}_{n}=\mathbb{C}^{2}$.
They can be represented by the elements of monodromy matrix as
\begin{align}
\rE_{n}^{ab}=\left\{ \prod_{k=1}^{n-1}\mathcal{T}(\theta_{k}+i/2)\right\} T^{ab}(\theta_{n}+i/2)\left\{ \prod_{k=1}^{n}\mathcal{T}(\theta_{k}+i/2)\right\} ^{-1}\label{eq:sQISP}
\end{align}
where
\begin{align}
T^{11}(u)=A(u),\quad T^{12}(u)=B(u),\quad T^{21}(u)=C(u),\quad T^{22}(u)=D(u).
\end{align}
Once inserted inside a correlator, the transfer matrices in (\ref{eq:sQISP})
act on a Bethe state and can be replaced by their eigenvalues
\begin{align}
\langle\mathbf{u}|\rE_{n}^{ab}|\mathbf{u}\rangle=\frac{\langle\mathbf{u}|T^{ab}(\theta_{n}+i/2)|\mathbf{u}\rangle}{t_{\mathbf{u}}(\theta_{n}+i/2)}\label{eq:lg1}
\end{align}
where $t_{\mathbf{u}}(u)$ is given by (\ref{trans}). At $u=\theta_{n}+i/2$,
we have
\begin{align}
t_{\mathbf{u}}(\theta_{n}+i/2)=Q_{\bm{\theta}}(\theta_{n}+i)\frac{Q_{\mathbf{u}}^{-}(\theta_{n})}{Q_{\mathbf{u}}^{+}(\theta_{n})}.
\end{align}
The generalization of (\ref{eq:lg1}) to a string of $l+1$ operators
is straightforward
\begin{align}
\langle\mathbf{u}|\rE_{n}^{a_{0}b_{0}}\cdots\rE_{n+l}^{a_{l}b_{l}}|\mathbf{u}\rangle=\left(\prod_{k=0}^{l}\frac{Q_{\mathbf{u}}^{+}(\theta_{n+k})}{Q_{\bm{\theta}}(\theta_{n+k}+i)Q_{\mathbf{u}}^{-}(\theta_{n+k})}\right)\langle\mathbf{u}|\prod_{k=0}^{l}T_{n+k}^{a_{k}b_{k}}(\theta_{n+k}+i/2)|\mathbf{u}\rangle\label{eq:lg}
\end{align}
We can now compute the r.h.s. of (\ref{eq:lg}) by using the Yangian
algebra and the Slavnov formula.

\section{Finite volume diagonal form factors}

The interest in the volume dependence of diagonal matrix elements
of local operators is due to their appearance in different physical
quantities. They are central objects in the calculation of finite
temperature one-point functions \cite{Leclair-Mussard,Pozsgay:2013jua},
as well as they play an essential role in the form factor perturbation
theory \cite{Delfino:1996xp}. The short distance behavior of correlation
functions also involves the vacuum expectation values as basic ingredients
\cite{Zamolodchikov:1990bk}. Recently, it was conjectured \cite{Bajnok:2014sza}
that the Heavy-Heavy-Light symmetric structure constants of the AdS/CFT
correspondence (\ref{C}) are also related to these quantities. This
latter case is of the central interest of the current paper. To better
understand this conjecture, in this section we summarize the
theory of form factors in integrable models, starting from the infinite
volume description and then take into account the finite volume
corrections, up to wrapping.

\subsection{Form factors in infinite volume}

We consider a $1+1$ dimensional integrable quantum field theory defined
by its $S$-matrix. For simplicity we restrict ourselves to diagonally
scattering theories with a self-conjugated particle, the generalization
to any diagonally scattering theory is straightforward. For a detailed
review of the general case see \cite{Smirnov:1992vz}.

The infinite volume states can be characterized by the set of momenta
of particles. In $1+1$ dimension it is convenient to label the particles
by their rapidities $u_{i}$, the energy $\varepsilon(u)$ and momentum
$p(u)$ being a single valued functions. In the remote past, an \emph{in}
state consists of particles of ordered rapidities: the fastest one
is the leftmost while the slowest is the rightmost. Contrarily, the
particles in an \emph{out} state are reversely ordered,
\begin{align}
\vert u_{1},\cdots,u_{N}\rangle=\begin{cases}
\vert u_{1},\cdots,u_{N}\rangle^{\text{in}} & \qquad u_{1}>\cdots>u_{N}\\
\vert u_{1},\cdots,u_{N}\rangle^{\text{out}} & \qquad u_{1}<\cdots<u_{N}
\end{cases}\label{eq:asym_states}
\end{align}
The infinite volume states that differ only in the order of rapidities
are related by the two particle $S$-matrix\footnote{Although the states with non-ordered rapidities are not physical.}
\begin{align}
\vert u_{1},\cdots,u_{i},u_{i+1},\cdots,u_{N}\rangle=S(u_{i},u_{i+1})\,\vert u_{1},\cdots,u_{i+1},u_{i},\cdots,u_{N}\rangle\label{eq:gen_state_inf_vol}
\end{align}
The energy of a multiparticle state is the sum of the one particle
energies
\begin{align}
E(u_{1},\cdots,u_{N})=\sum_{i=1}^{N}\varepsilon(u_{i}).
\end{align}
In infinite volume we normalize the \emph{in} states as
\begin{align}
\phantom{I}^{\text{in}}\langle u'_{1},\cdots,u'_{M}\vert u_{1},\cdots,u_{N}\rangle^{\text{in}}=(2\pi)^{N}\delta_{NM}\,\delta(u_{1}-u'_{1})\cdots\delta(u_{N}-u'_{N}),\label{eq:Inf_Vol_norm}
\end{align}
and the norm of a general state can be determined from (\ref{eq:Inf_Vol_norm})
by (\ref{eq:gen_state_inf_vol}).

Let us consider the matrix elements of a \emph{local} operator $\mathcal{O}(t,x)$
between asymptotic states. The space-time dependence can be easily
factored out
\begin{align*}
\phantom{I}^{\text{out}}\left\langle u'_{1},\cdots,u'_{M}\right|\mathcal{O}(x,t)\left|u_{1},\cdots,u_{N}\right\rangle ^{\text{in}}=e^{it\Delta E-ix\Delta P}\phantom{I}^{\text{out}}\left\langle u'_{1},\cdots,u'_{M}\right|\mathcal{O}(0,0)\left|u_{1},\cdots,u_{N}\right\rangle ^{\text{in}},
\end{align*}
where
\begin{align}
\Delta E= & \,\sum_{j=1}^{M}\varepsilon(u'_{j})-\sum_{k=1}^{N}\varepsilon(u_{k}),\\
\Delta P= & \,\sum_{j=1}^{M}p(u'_{j})-\sum_{k=1}^{N}p(u_{k}),\nonumber
\end{align}
and we define the \emph{form factor} of operator $\mathcal{O}$ as
\begin{equation}
F_{M,N}^{\mathcal{O}}\left(u'_{1},\cdots,u'_{M}\vert u_{1},\cdots,u_{N}\right)=\phantom{I}^{\text{out}}\left\langle u'_{1},\cdots,u'_{M}\right|\mathcal{O}(0,0)\left|u_{1},\cdots,u_{N}\right\rangle ^{\text{in}}
\end{equation}
The form factors are \emph{a priori} defined for ordered set of incoming
and outgoing rapidities but can be analytically continued by (\ref{eq:gen_state_inf_vol}).
A form factor is a meromorphic function in all variables and each
pole has a physical origin \cite{Smirnov:1992vz}.

Suppose that the theory possesses crossing symmetry, i.e. a transformation
which maps an outgoing particle with rapidity $u$ to an incoming
anti-particle with rapidity $\bar{u}$. The crossing symmetry implies the crossing equation
for the form factors which, in case of a single self-conjugated particle,
reads as
\begin{align}
 & \, F_{M,N}^{\mathcal{O}}\left(u'_{1},\cdots,u'_{M}\vert u_{1},\cdots,u_{N}\right)=F_{M-1,N+1}^{\mathcal{O}}\left(u'_{1},\cdots,u'_{M-1}\vert\bar{u}_{M},u_{1},\cdots,u_{N}\right)\label{eq:FF_crossing}\\
 & \qquad+\sum_{k=1}^{N}\left\langle u'_{M}\vert u_{k}\right\rangle \prod_{l=1}^{k-1}S\left(u_{l},u_{k}\right)F_{M-1,N-1}^{\mathcal{O}}\left(u'_{1},\cdots,u'_{M-1}\vert u_{1},\cdots,\hat{u}_{k},\cdots,u_{N}\right)\nonumber
\end{align}
where the terms on the second line of (\ref{eq:FF_crossing}) describe
disconnected processes that occur if one of the incoming and outgoing
particle has the same rapidity. The hat $\hat{u}_{k}$ denotes that
$u_{k}$ is missing from the list of rapidities. By using the crossing
relation all form factors can be expressed in terms of \emph{elementary}
form factors
\begin{align}
F_{N}^{\mathcal{O}}\left(u_{1},\cdots,u_{N}\right)=\langle0\vert\mathcal{O}(0,0)\vert u_{1},\cdots,u_{N}\rangle.
\end{align}
These elementary form factors satisfy several functional relations,
called the form factor axioms. As these axioms are not relevant for
the aims of this paper we will not list them here but rather refer
to \cite{Smirnov:1992vz} for relativistic models and \cite{Klose:FF}
for the AdS/CFT case.

\subsubsection{Diagonal form factors}

The diagonal form factor of the local operator $\mathcal{O}$, defined
as
\begin{align}
\phantom{I}^{\text{in}}\langle u_{1},\cdots,u_{N}\vert\mathcal{O}(0,0)\vert u_{1},\cdots,u_{N}\rangle^{\text{in}},\label{eq:diag_FF}
\end{align}
is singular due to the disconnected terms in the crossing relation
(\ref{eq:FF_crossing}). To avoid the singularities we regularize
it by slightly shifting the outgoing rapidities. After crossing we
get
\begin{align}
F_{2N}^{\mathcal{O}}\left(\bar{u}_{1}+\epsilon_{1},\cdots,\bar{u}_{N}+\epsilon_{N},u_{N},\cdots,u_{1}\right)=\langle0\vert\mathcal{O}\vert\bar{u}_{1}+\epsilon_{1},\cdots,\bar{u}_{N}+\epsilon_{1},u_{1},\cdots,u_{N}\rangle^{\text{in}}\label{eq:diag_FF_regul}
\end{align}
The diagonal limit, $\epsilon_{i}\rightarrow0$, of (\ref{eq:diag_FF_regul})
is not well-defined. It was first noticed in \cite{Delfino:1996xp}
that the singular parts vanish in the limit when all $\epsilon_{i}\rightarrow0$,
but the result depends on the direction of the limit. Its general
structure can be written as
\begin{align}
 & \, F_{2N}^{\mathcal{O}}\left(\bar{u}_{1}+\epsilon_{1},\cdots,\bar{u}_{N}+\epsilon_{N},u_{N},\cdots,u_{1}\right)\label{eq:Reg_DiagFF_struct}\\
= & \,\prod_{i=1}^{N}\frac{1}{\epsilon_{i}}\cdot\sum_{i_{1}=1}^{N}\sum_{i_{2}=1}^{N}\cdots\sum_{i_{N}=1}^{N}a_{i_{1}i_{2}\cdots i_{N}}(u_{1},\cdots,u_{N})\,\epsilon_{i_{1}}\epsilon_{i_{2}}\cdots\epsilon_{i_{N}}+\cdots\nonumber
\end{align}
where $a_{i_{1}i_{2}...i_{N}}$ is a completely symmetric tensor of
rank $N$. The ellipsis denote terms which vanish in the $\epsilon_{i}\rightarrow0$
limit.

There are two generally used regularization scheme in the literature.
The first is the so-called \emph{symmetric} evaluation when we set
all $\epsilon_{i}$ to be the same
\begin{equation}
F_{2N}^{\mathcal{O},s}\left(u_{1},\cdots,u_{N}\right)=\lim_{\epsilon\rightarrow0}F_{2N}^{\mathcal{O}}\left(\bar{u}_{1}+\epsilon,\cdots,\bar{u}_{N}+\epsilon,u_{N},\cdots,u_{1}\right).\label{eq:symmetric_FF}
\end{equation}
The second scheme is called \emph{connected}, in which the diagonal
form factors are defined as the finite part of (\ref{eq:Reg_DiagFF_struct}),
i.e. the $\epsilon$-independent term,
\begin{align}
F_{2N}^{\mathcal{O},c}\left(u_{1},\cdots,u_{N}\right)=N!\, a_{12\cdots N}\ .\label{eq:connected_FF}
\end{align}
Both the symmetric and the connected diagonal form factors are completely
symmetric in the rapidity variables $u_{1},\cdots,u_{N}$. Of course
these two quantities are not independent and each can be expressed
with use of the other \cite{Pozsgay:2007gx}.

\subsection{Diagonal form factors in finite volume}

In this section we will summarize the results about the volume dependence
of the diagonal form factors in all polynomial orders in the inverse
of the volume, neglecting the exponentially small wrapping corrections,
following \cite{Pozsgay:2007gx,Pozsgay:2007kn}.

In finite volume $L$, the rapidities are quantized and a generic
multiparticle state can be labeled by the Bethe quantum numbers $\vert\{I_{1},\cdots,I_{N}\}\rangle_{L}$.
In finite volume we cannot order the particles by spatial separation
in the remote past or future, as we did in the infinite volume case
(\ref{eq:gen_state_inf_vol}). In finite volume the states are completely
symmetric under the exchange of particles and can be characterized
by the \emph{set} of quantum numbers. We adapt our notation to the
conventions used in \cite{Pozsgay:2007gx,Pozsgay:2007kn} and order
the quantum numbers in a monotonly decreasing sequence, $I_{1}\geq\cdots\geq I_{N}$
\footnote{Apart from the free boson case all known $S$-matrix obey the property
$S(u,u)=-1$ which is an effective Pauli exclusion. In this cases
we have $I_{1}>\cdots>I_{N}$. %
}. The quantized rapidities with the quantum numbers $\{I_{1},\cdots,I_{N}\}$
are solutions of the corresponding Bethe Ansatz equations. Similarly
to (\ref{eq:Phik}) we define
\begin{align}
\Phi_{j}(\{u_{1},\cdots,u_{N}\})=p(u_{j})L-i\sum_{{k=1\atop k\ne j}}\log S(u_{k},u_{j})\,,\label{eq:BAE}
\end{align}
and the quantization condition reads as
\begin{align}
\Phi_{j}(\{u_{1},\cdots,u_{N}\})=2\pi I_{j},\qquad j=1,\cdots,N.
\end{align}
These finite volume states are orthogonal to each other
\begin{align}
\phantom{I}_{L}\langle\{J_{1},\cdots,,J_{M}\}\vert\{I_{1},\cdots,I_{N}\}\rangle_{L}\propto\delta_{NM}\,\delta_{I_{1},J_{1}}\cdots\delta_{I_{N},J_{N}}\label{eq:finite_vol_orthogon}
\end{align}
and their normalization is a question of convention.

One can change from the quantum number representation of states to
the rapidity representation which gives the direct connection between
the finite and infinite volume states \cite{Pozsgay:2007kn}. This
change of variables involves the Jacobian, which is the density of
$N$-particle states, defined as
\begin{align}
\varrho_{N}(u_{1},\cdots,u_{N})_{L}= & \,\det\mathcal{J}^{(N)}(u_{1},\cdots,u_{N})_{L}\label{eq:BetheYang_Jacobian}\\
\mathcal{J}_{k,l}^{(N)}(u_{1},\cdots,u_{N})_{L}= & \,\frac{\partial\Phi_{k}(u_{1},\cdots,u_{N})}{\partial u_{l}}\quad,\quad k,l=1,\cdots,N\,.\nonumber
\end{align}
where we explicitly indicated the volume dependence of these quantities.
The determinant (\ref{eq:BetheYang_Jacobian}) is closely related
to the Gaudin norm of Bethe states (\ref{eq:norm})%
\footnote{The Gaudin norm itself is not physical as it depends on the conventions.
However, in any convention, it is proportional to the Jacobian (\ref{eq:BetheYang_Jacobian}). %
} Then the relation between the infinite and finite volume states reads
as
\begin{align}
\vert\{I_{1},\cdots,I_{N}\}\rangle_{L}=\frac{1}{\sqrt{\varrho_{N}(u_{1},\cdots,u_{N})_{L}\prod_{i<j}S(u_{i},u_{j})}}\vert u_{1},\cdots,u_{N}\rangle\label{eq:Finite_vol_state}
\end{align}
where the rapidities $\{u_{i}\}$ are the solutions of the Bethe Ansatz
equations (\ref{eq:BAE}) corresponding to the quantum numbers $\{I_{1},\cdots,I_{N}\}$.
This identification holds up to exponential corrections. The product
of $S$-matrices in the denominator ensures that the finite volume
state is indeed symmetric under the exchange of particles.

Defining the system in finite volume regularizes all the divergences
appearing in the diagonal limit of form factors (\ref{eq:Reg_DiagFF_struct}),
thus the normalized finite volume diagonal matrix element
\begin{align}
\frac{\phantom{I}_{L}\langle\{I_{1},\cdots,I_{N}\}\vert\mathcal{O}(0,0)\vert\{I_{1},\cdots,I_{N}\}\rangle_{L}}{\phantom{I}_{L}\langle\{I_{1},\cdots,I_{N}\}\vert\{I_{1},\cdots,I_{N}\}\rangle_{L}}\label{eq:Fin_Vol_diag_FF}
\end{align}
is finite, completely well defined and does not depend on the normalization
of states. However, it is a challenging task to relate them to the
infinite volume ones in the general case \cite{Leclair-Mussard,Pozsgay:2013jua}.
The problem become considerably simpler if we neglect the exponentially
small wrapping corrections.

Up to wrapping, the finite volume $N$-particle diagonal form factor
(\ref{eq:Fin_Vol_diag_FF}) can be expressed as a sum over the bipartite
partitions of the full set $\{1,2,\cdots,N\}$, involving the infinite
volume form factors and some kind of densities of states. As the diagonal
form factors in infinite volume depend on the regularization scheme,
this series is also scheme dependent. In case of the connected evaluation
the relation reads as \cite{Saleur:commentFF,Pozsgay:2007gx}
\begin{equation}
\frac{\phantom{I}_{L}\langle\{I_{1},\cdots,I_{N}\}\vert\mathcal{O}(0,0)\vert\{I_{1},\cdots,I_{N}\}\rangle_{L}}{\phantom{I}_{L}\langle\{I_{1},\cdots,I_{N}\}\vert\{I_{1},\cdots,I_{N}\}\rangle_{L}}=\frac{1}{\rho_{N}(\{1,\cdots,N\})}\sum_{\alpha\subseteq\{1,\dots,N\}}f^{\mathcal{O}}\left(\{u_{k}\}_{k\in\bar{\alpha}}\right)\rho_{N}\left(\alpha\right)\label{eq:Connected_expansion}
\end{equation}
where $\bar{\alpha}$ denotes the complement of $\alpha$ in the full
set. The functions appearing on the right hand side are exactly the
connected diagonal form factors
\begin{align}
f^{\mathcal{O}}(u_{1},\cdots,u_{l})=F_{2l}^{\mathcal{O},c}(u_{1},\cdots,u_{l})
\end{align}
The functions $\rho_{N}$ are defined as the diagonal minor determinants
of the $N$-particle Jacobian (\ref{eq:BetheYang_Jacobian}),
\begin{equation}
\rho_{N}(\alpha)=\det_{k,l\in\alpha}\mathcal{J}_{k,l}^{(N)}(u_{1},\cdots,u_{N})_{L}\quad,\qquad\alpha\subseteq\{1,\cdots,N\}.\label{eq:Diag_minor_GN}
\end{equation}
They can also be referred to as partial Gaudin norms. As special cases
we have
\begin{equation}
\rho_{N}(\{1,\cdots,N\})=\varrho_{N}(u_{1},\cdots,u_{N})_{L}\quad;\qquad\rho_{N}(\emptyset)=1.\label{eq:Rewritten_BY_Jacobian}
\end{equation}
We want to emphasize that the function $\rho_{N}(\alpha)$ depend
on \emph{all} the $N$ rapidities. The set of rapidities $\{u_{i}\}$
in the right hand side of (\ref{eq:Connected_expansion}) is the solution
of the Bethe Ansatz equations (\ref{eq:BAE}) corresponding to the
quantum numbers $\{I_{1},\cdots,I_{N}\}$. Thus, the \emph{explicit}
volume dependence is encoded \emph{only} into the factors $\rho_{N}$,
the connected form factors $f^{\mathcal{O}}$ depend on the volume
only \emph{implicitly} via the Bethe Ansatz equations.

As the connected and symmetric diagonal form factors are not independent,
we can express the finite volume matrix element in the symmetric regularization
scheme. In this case the series take the form \cite{Pozsgay:2007kn}
\begin{equation}
\frac{\phantom{I}_{L}\langle\{I_{1},\cdots,I_{N}\}\vert\mathcal{O}(0,0)\vert\{I_{1},\cdots,I_{N}\}\rangle_{L}}{\phantom{I}_{L}\langle\{I_{1},\cdots,I_{N}\}\vert\{I_{1},\cdots,I_{N}\}\rangle_{L}}=\frac{1}{\rho_{N}(\{1,\cdots,N\})}\sum_{\alpha\subseteq\{1,\dots,n\}}F_{2\left|\bar{\alpha}\right|}^{s}\left(\{u_{k}\}_{k\in\bar{\alpha}}\right)\rho_{\left|\alpha\right|}\left(\alpha\right).\label{eq:Symmetric_expansion}
\end{equation}
Here again, the rapidities $\{u_{i}\}$ are the solutions of the Bethe
Ansatz equations (\ref{eq:BAE}) with the quantum numbers $\{I_{1},\cdots,I_{N}\}$.
The $\rho_{\left|\alpha\right|}$ functions appearing in the sum are
the $\left|\alpha\right|$-particle densities of state (\ref{eq:Rewritten_BY_Jacobian},\ref{eq:BetheYang_Jacobian})
evaluated at the rapidities $\{u_{i}\}_{i\in\alpha}$. Note that,
contrary to the connected expansion, they depend \emph{only} on the
rapidities labeled by the set $\alpha$. The explicit volume dependence
is carried only by the $\rho$ functions.

\subsubsection{Form factor of densities of conserved charges}

An important special case of local operators is the density of a conserved
quantity,
\[
Q=\int_{0}^{L}J(x,t)dx
\]
where $Q$ acts diagonally and additively on the multiparticle states.
Its density therefore satisfies
\[
\frac{\phantom{I}_{L}\langle\{I_{1},\cdots,I_{N}\}\vert J(0,0)\vert\{I_{1},\cdots,I_{N}\}\rangle_{L}}{\phantom{I}_{L}\langle\{I_{1},\cdots,I_{N}\}\vert\{I_{1},\cdots,I_{N}\}\rangle_{L}}=\frac{1}{L}\sum_{j=1}^{N}q(u_{j}),
\]
where $\{u_{i}\}$ are the solutions of the Bethe Ansatz equations
(\ref{eq:BAE}) corresponding to the quantum numbers $\{I_{1},\cdots,I_{N}\}$,
and $q(u)$ is the one-particle eigenvalue of the operator $Q$.

A compact expression for the connected diagonal form factors of these
densities was presented in \cite{Leclair-Mussard,Saleur:commentFF},
however the proof was found recently \cite{Janos}.
The connected form factors can be cast into the form
\begin{equation}
F_{2N}^{J,c}(u_{1}\cdots,u_{N})=\sum_{\sigma\in S_{N}}p'(u_{\sigma(1)})\varphi(u_{\sigma(1)},u_{\sigma(2)})\varphi(u_{\sigma(2)},u_{\sigma(3)})\cdots\varphi(u_{\sigma(N-1)},u_{\sigma(N)})q(u_{\sigma(N)})\label{eq:FF_conn_conservedQ}
\end{equation}
where the summation runs over all the permutation of the set $\{1,\cdots,N\}$.
Here, $p'$ denotes the derivative of the momentum w.r.t the rapidity,
\[
p'(u)=\frac{\partial}{\partial u}p(u).
\]
For massive relativistic models we have $p(u)=m\sinh u$ and $\varepsilon(u)=m\cosh u$,
so that $p'(u)=\varepsilon(u)$ and (\ref{eq:FF_conn_conservedQ})
reduces to the expression presented in \cite{Leclair-Mussard,Saleur:commentFF}.
However, in the case of the Heisenberg XXX$_{1/2}$ spin chain an
extra sign appears, $p'(u)=-\varepsilon(u)$.

\subsection{Conjecture for the symmetric structure constants}

Based on explicit calculations the authors of \cite{Bajnok:2014sza}
conjectured that the Heavy-Heavy-Light symmetric structure constant in the $\mathfrak{su}(2)$ sector
of ${\cal N}=4$ SYM is equal to a finite volume form factor and its volume dependence (up to wrapping corrections)
at any coupling has the form \ur{fvf}.
Let us suppose that the two, conjugated heavy operators correspond
to a multiparticle state, labeled by the rapidities $\{u_{i}\}$,
in finite, but large volume $L$, such that the exponential corrections
are negligible. Then the symmetric structure constant are conjectured
to be the finite volume diagonal form factor of the vertex operator
$\mathcal{O}$ of the light operator,
\begin{equation}
C_{\alpha}=\phantom{I}_{L}\langle u_{1},\cdots,u_{N}\vert\mathcal{O}\vert u_{1},\cdots,u_{N}\rangle_{L}\label{eq:BJW_conj}
\end{equation}
that can be expressed in terms of the infinite volume quantities (\ref{eq:Connected_expansion})
and (\ref{eq:Symmetric_expansion}).

\section{Matrix elements of spin operators}

\label{sec:HeisenbergFF} In this section we study the diagonal matrix
elements of spin operators of the Heisenberg spin chain using the
ABA and the solution of QISP discussed in section \ref{sec:ABA}.
We show that, in general, the matrix elements can be written as linear
combinations of a special kind of Slavnov determinant. In section \ref{sec:FVE}, we show that
this determinant has the structure conjectured in \cite{Bajnok:2014sza},
namely it can be written as linear combinations of diagonal minors
of Gaudin determinants (\ref{eq:Connected_expansion}). We call the procedure
of expanding quantities in terms of diagonal minors of Gaudin determinants
the \emph{finite volume expansion}, as it captures all the finite volume dependence. We will discuss the case $l_{s}=1,2$
in detail and comment on the general $l_{s}>2$ case.

\subsection{Form factors of length-$2$ operators}

We have shown in section \ref{eq:CFtoFF} that all the length-$2$
diagonal matrix elements can be written as linear combinations of
the following building blocks
\begin{align}
\cF^{o_{1}}=\langle\mathbf{u}|o_{1}(n)|\mathbf{u}\rangle,\qquad\cF^{o_{2}^{i}}=\langle\mathbf{u}|o_{2}^{i}(n)|\mathbf{u}\rangle,\quad i=1,...,3\label{blockF}
\end{align}
where the local operators $o_{1}(n)$ and $o_{2}^{i}(n)$ are given
in (\ref{eq:o1}) and (\ref{eq:o2}). According to (\ref{eq:lg}),
these matrix elements are proportional to the following quantities
\begin{align}
\cF^{o_{1}}\propto F^{A}\equiv\langle\mathbf{u}|A(\theta_{n}+i/2)|\mathbf{u}\rangle.\label{eq:block1}
\end{align}
\begin{align}
\cF^{o_{2}^{1}}\propto F^{AA}\equiv & \,\langle\mathbf{u}|A(\theta_{n}+i/2)A(\theta_{n+1}+i/2)|\mathbf{u}\rangle,\label{eq:block2}\\
\cF^{o_{2}^{2}}\propto F^{BC}\equiv & \,\langle\mathbf{u}|B(\theta_{n}+i/2)C(\theta_{n+1}+i/2)|\mathbf{u}\rangle,\nonumber \\
\cF^{o_{2}^{3}}\propto F^{CB}\equiv & \,\langle\mathbf{u}|C(\theta_{n}+i/2)B(\theta_{n+1}+i/2)|\mathbf{u}\rangle,\nonumber \\
\cF^{o_{2}^{4}}\propto F^{DD}\equiv & \,\langle\mathbf{u}|D(\theta_{n}+i/2)D(\theta_{n+1}+i/2)|\mathbf{u}\rangle.\nonumber
\end{align}

In order to compute the building blocks (\ref{eq:block1}) and (\ref{eq:block1}),
we act all the operators on the ket state $|\mathbf{u}\rangle$. The
action of $A$ and $D$ on a Bethe state is
\begin{align}
 & A(v)|\mathbf{u}\rangle=a(v)\frac{Q_{\mathbf{u}}(v-i)}{Q_{\mathbf{u}}(v)}|\mathbf{u}\rangle+\sum_{n=1}^{N}\rM_{n}(v)\,|\{\mathbf{u},v\}\setminus\{u_{n}\}\rangle,\label{eq:actAD}\\
 & D(v)|\mathbf{u}\rangle=d(v)\frac{Q_{\mathbf{u}}(v+i)}{Q_{\mathbf{u}}(v)}|\mathbf{u}\rangle+\sum_{n=1}^{N}\rN_{n}(v)\,|\{\mathbf{u},v\}\setminus\{u_{n}\}\rangle\nonumber
\end{align}
where $\rM_{k}(v)$, $\rN_{k}(v)$ are given by
\begin{align}
 & \rM_{k}(v)=\frac{ia(u_{n})}{v-u_{n}}\prod_{j\ne n}^{N}\frac{u_{n}-u_{j}-i}{u_{n}-u_{j}},\\
 & \rN_{k}(v)=\frac{id(u_{n})}{u_{n}-v}\prod_{j\ne n}^{N}\frac{u_{n}-u_{j}+i}{u_{n}-u_{j}}.\nonumber
\end{align}
These relations can be derived from the Yangian algebra (\ref{eq:Yangian}).
From (\ref{eq:actAD}) we see that the action of the operators $A$
and $D$ on a Bethe state preserve the number of magnons. In addition
to the original Bethe state $|\mathbf{u}\rangle$, there is a sum
of Bethe states $|\{\mathbf{u},v\}\setminus\{u_{k}\}\rangle$ where
one of the rapidities $u_{k}$ is replaced by the spectral parameter
$v$ of the operator. These are called the \emph{unwanted terms} and
are off-shell for generic $v$. On the other hand, they are not too
far from the on-shell Bethe state $|\mathbf{u}\rangle$ since most
of the rapidities remain unchanged.

The action of the $C$ operator on the Bethe state is more involved
\begin{align}
C(v)|\mathbf{u}\rangle=\sum_{n=1}^{N}\rK_{n}\,|\{\mathbf{u}\}\setminus\{u_{n}\}\rangle+\sum_{k>n}\rK_{kn}\,|\{\mathbf{u},v\}\setminus\{u_{k},u_{n}\}\rangle\label{eq:actC}
\end{align}
where
\begin{align}
\rK_{n}= & \,\frac{ia(v)d(u_{n})}{u_{n}-v}\prod_{j\ne n}^{N}\frac{u_{j}-u_{n}-i}{u_{j}-u_{n}}\cdot\frac{u_{j}-v+i}{u_{j}-v}+\\
 & \,\frac{ia(u_{n})d(v)}{v-u_{n}}\prod_{j\ne n}^{N}\frac{u_{j}-u_{n}+i}{u_{j}-u_{n}}\cdot\frac{u_{j}-v-i}{u_{j}-v}\nonumber \\
\rK_{kn}= & \,\frac{d(u_{k})a(u_{n})}{(u_{k}-v)(u_{n}-v)}\frac{u_{k}-u_{n}+i}{u_{k}-u_{n}}\prod_{j\ne k,n}\frac{u_{j}-u_{k}-i}{u_{j}-u_{k}}\cdot\frac{u_{j}-u_{n}+i}{u_{j}-u_{n}}+\nonumber \\
 & \,\frac{d(u_{n})a(u_{k})}{(u_{n}-v)(u_{k}-v)}\frac{u_{k}-u_{n}-i}{u_{k}-u_{n}}\prod_{j\ne k,n}\frac{u_{j}-u_{k}+i}{u_{j}-u_{k}}\cdot\frac{u_{j}-u_{n}-i}{u_{j}-u_{n}}\nonumber
\end{align}
The coefficients $\rK_{n}$ and $\rK_{kn}$ can be expressed in terms
of $\rM_{n}$ and $\rN_{n}$
\begin{align}
\rK_{n}(v)= & \,\rM_{n}(v)\rN_{0}(v)\frac{v-u_{n}}{v-u_{n}+i}+\rM_{0}(v)\rN_{n}(v)\frac{v-u_{n}}{v-u_{n}-i},\\
\rK_{nk}(v)= & \,\rM_{k}(v)\rN_{n}(v)\frac{u_{n}-u_{k}}{u_{n}-u_{k}+i}+\rM_{n}(v)\rN_{k}(v)\frac{u_{n}-u_{k}}{u_{n}-u_{k}-i}\nonumber
\end{align}
if we define
\begin{align}
\rM_{0}\equiv a(v)\frac{Q_{\mathbf{u}}(v-i)}{Q_{\mathbf{u}}(v)},\qquad\rN_{0}\equiv d(v)\frac{Q_{\mathbf{u}}(v+i)}{Q_{\mathbf{u}}(v)}.
\end{align}
From (\ref{eq:actC}) it is clear that $C$ reduces the number of
magnons by one. For diagonal matrix elements, any $C$ operator has
to be accompanied by a $B$ operator in order to preserve $S^{z}$,
and obtain non-vanishing results. For length-$2$ operators the only
possibilities are $B(u)C(v)$ and $C(u)B(v)$. Both combinations preserve
the number of magnons but will lead to a sum of unwanted terms with
one or two magnons replaced by the spectral parameters of the operators.

\subsection{Form factors of length $l_{s}$ operators}
In general, the action of $l_{s}$ ABA operators on a Bethe state
generates unwanted terms with at most $l_{s}$ rapidities replaced
by the spectral parameters of the operators. In particular, it is
clear now that all the building blocks (\ref{eq:block1}) and (\ref{eq:block2})
can be written as the scalar products of the following three types
\begin{align}
\langle\mathbf{u}|\mathbf{u}\rangle,\quad\langle\mathbf{u}|\{\mathbf{u},\theta_{n}^{+}\}\setminus\{u_{k}\}\rangle,\quad\langle\mathbf{u}|\{\mathbf{u},\theta_{n}^{+},\theta_{n+1}^{+}\}\setminus\{u_{j},u_{k}\}\rangle.
\end{align}
where we have used the notation $\theta_{n}^{+}=\theta_{n}+i/2$.
The first scalar product is the Gaudin norm (\ref{eq:norm}). The finite volume dependence of
other two determinants is the subject of discussion in section \ref{sec:FVE}.\par

The discussion of the previous subsection can be generalized to the
diagonal matrix elements of the operators with $l_{s}>2$. As before,
any such matrix element can be spanned by some building blocks such
as $F^{A\cdots A}$, $F^{BCA\cdots A}$, $F^{D\cdots D}$. Of course,
the number of the building blocks grows with the length of the operator.

We act all the ABA operators on the ket state which give rise to the
unwanted terms
\begin{align}
|\{\mathbf{u},\theta_{n}^{+},...,\theta_{n+M}^{+}\}\setminus\{u_{k_{1}},...,u_{k_{M}}\}\rangle,\qquad M\le l_{s}.
\end{align}
Thus the diagonal matrix element of any length-$l_{s}$ operator can be
written as a linear combination of the following scalar products
\begin{align}
\langle\mathbf{u}|\{\mathbf{u},\theta_{n}^{+},...,\theta_{n+M}^{+}\}\setminus\{u_{k_{1}},...,u_{k_{M}}\}\rangle,\quad M\le l_{s}.\label{eq:mGN}
\end{align}
The number of terms and the complexity of the coefficients will grow
quickly with the increase of number of magnons and length of the operators,
nevertheless the structure is robust.

\section{Finite volume expansion}

\label{sec:FVE} In this section, we analyze the structure of the
scalar products (\ref{eq:mGN}) and show that any of them can
be expanded in terms of diagonal minors of Gaudin norms. We call this
procedure the \emph{finite volume expansion}.

Above mentioned scalar products can be computed by the Slavnov determinant formula. In
the Slavnov determinant (\ref{sp}), the denominator is a simple Cauchy
determinant and can be computed readily. We therefore focus on the
non-trivial numerator $\det_{jk}\Omega(u_{j},v_{k})$. Let us first
consider the scalar product for the case of length-$2$ operators, $\langle\mathbf{u}|\{\mathbf{u},\theta_{n},\theta_{n+1}\}\setminus\{u_{j},u_{k}\}\rangle$.
The determinant takes the following form
\begin{align}
\det\Omega=\left|\begin{array}{ccccccc}
i\,\phi_{11} & \cdots & \bOmega{1j} & \cdots & \bOmega{1k} & \cdots & i\,\phi_{1N}\\
i\,\phi_{21} & \cdots & \bOmega{2j} & \cdots & \bOmega{2k} & \cdots & i\,\phi_{2N}\\
\vdots & \ddots & \vdots & \ddots & \vdots & \ddots & \vdots\\
i\,\phi_{N1} & \cdots & \bOmega{Nj} & \cdots & \bOmega{Nk} & \cdots & i\,\phi_{NN}
\end{array}\right|.
\end{align}
where we have defined $i\,\phi_{jk}=\Omega(u_{j},u_{k})$ and $\Omega_{ik}=\Omega_{ik}(u_{i},\theta_{n}+i/2)$.
The procedure is straightforward: \emph{perform Laplace expansion
with respect to the column or row that does not have any element of
the form $\phi_{nn}$ repeatedly, until one can not do it further}.
Note that after one Laplace expansion, we will obtain sub-determinants.
We shall also perform the same procedure for all the sub-determinants
until it terminates. This procedure will terminate when all the determinants
in the expression take the form of diagonal minors (\ref{eq:Diag_minor_GN})
of Gaudin norm (\ref{eq:BetheYang_Jacobian})
\begin{align}
\rho_{N}(\{i_{1},\cdots,i_{m}\})=(-1)^m\left|\begin{array}{cccc}
\,\phi_{i_{1}i_{1}} & \cdots & \cdots & \cdots\\
\cdots & \,\phi_{i_{2}i_{2}} & \cdots & \cdots\\
\vdots & \vdots & \ddots & \vdots\\
\cdots & \cdots & \cdots & \,\phi_{i_{m}i_{m}}
\end{array}\right|
\end{align}
Therefore, the following expansion holds
\begin{align}
\langle\mathbf{u}|\{\mathbf{u},\theta_{n},\theta_{n+1}\}\setminus\{u_{j},u_{k}\}\rangle=\sum_{\alpha\subseteq A}\cF(\bar{\alpha})\,\rho_{N}(\alpha),
\end{align}
where $A=\{1,...,\hat{j},...,\hat{k},...,N\}$%
\footnote{Here $\hat{j}$ and $\hat{k}$ mean these two indices are absent.%
} and the summation runs over all possible subsets $\alpha$ of $A$.
Here $\bar{\alpha}$ is the complement of $\alpha$ in $A$. For an
explicit and simple example, see Appendix \ref{Appendix1}.

Finally we need to justify why we call this procedure ``finite volume
expansion''. From a simple analysis below, it is clear that all the
explicit $L$ dependence are contained in the diagonal minors of the
Gaudin norm. In the ABA approach, the diagonal matrix elements are
given in terms of the following functions: the eigenvalue of the diagonal
elements of the transfer matrix, $a(u)$ and $d(u)$, the products
of functions $f(u-v)$ and $g(u-v)$ (\ref{eq:actAD},\ref{eq:actC}),
and the matrix elements in the Slavnov determinant formula $\Omega_{jk}$
and $\phi_{jk}$. Under proper normalization, the functions $a(u)$
and $d(u)$ always appear in the expression as the ratio $a(u)/d(u)=e^{ipL}$.
In fact, this kind of phase factor is either canceled by the same
factors from the norm, or be replaced by products of scattering matrices
using the Bethe Ansatz equations and they do not appear in the final
expression. The products of $f(u-v)$ and $g(u-v)$ functions do not
depend on $L$. The matrix element $\Omega_{jk}$ defined in (\ref{Slavnov})
also has no dependence on $L$. Finally, $\phi_{jk}$ with $j\ne k$
reads
\begin{align}
\phi_{jk}=\phi(u_{j},u_{k})=\frac{2}{(u_{j}-u_{k})^{2}+1},\quad j\neq k,\label{eq:Phijk}
\end{align}
again, do not depend on $L$. The only dependence on $L$ is hidden
in the diagonal element $\phi_{nn}$. Recall that we have
\begin{align}
\label{phinn}
\phi_{nn}=\sum_{m=1}^{L}\frac{1}{(u_{n}-\theta_{m})^{2}+1/4}-\sum_{{l=1\atop l\neq n}}^{N}\phi_{nl}
\end{align}
In the homogeneous limit, where $\theta_{m}=0$ ($m=1,\cdots,L$),
the first term of (\ref{phinn}) becomes $L/(u_{n}^{2}+1/4)$ which
depends linearly on $L$. When we perform the Laplace expansion, we carefully avoid expansion with this kind
of terms and they only appear in the diagonal minor $\rho_{N}(\alpha)$.
Therefore, the finite volume corrections are all contained in $\rho_{N}(\alpha)$.
This is one part of the conjecture in \cite{Bajnok:2014sza}%
\footnote{There the authors used an equivalent description of the diagonal matrix
element (\ref{eq:BJW_conj}) in terms of the symmetric expansion (\ref{eq:Symmetric_expansion}),
instead the connected one (\ref{eq:Connected_expansion}) that we
used here.%
}.

We have shown in section \ref{sec:HeisenbergFF} that any diagonal
matrix element can be written as a linear combination of specific determinant
with coefficients that do not depend explicitly on $L$.
As we showed above, these determinants allow finite volume expansion,
thus we can perform the finite volume expansion of any diagonal
matrix element in the Heisenberg spin chain. As was shown in section
\ref{eq:CFtoFF}, the diagonal matrix elements correspond to three-point
functions of HHL type. Therefore we have shown that the structure
of finite volume dependence of three-point functions conjectured in
\cite{Bajnok:2014sza} is also valid at weak coupling at the leading
order in the $\mathfrak{su}(2)$ sector. In section \ref{sec:loopFF},
we will show that the structure also holds at one-loop level.

This is only half of the story. In the conjecture \cite{Bajnok:2014sza},
each coefficient $\cF(\bar{\alpha})$ of $\rho(\alpha)$ is identified
with the form factor of the same operator in \emph{infinite volume}.
In order to check this statement, it is desirable to have a formulation
of the diagonal matrix elements of the Heisenberg spin chain directly
in infinite volume. However, we are not aware of such a formulation,
although it seems possible to do it in the framework of coordinate
Bethe Ansatz. In principle, the infinite volume form factors for our
case can also be obtained by first solving the Klose-McLoughlin axioms
\cite{Klose:FF} and then take the weak coupling limit. However,
no solution has been found up to now. Because of these reasons, we
are not able to confirm that the coefficients we obtain from finite
volume expansion are indeed the infinite volume form factors.

It is still of interest to know the explicit form of coefficients
from our finite volume expansion. These will be our predictions for
the diagonal form factors in the infinite volume theory. We perform
the finite volume expansion for all the diagonal matrix elements of
length-$1$ and length-$2$ operators and extract the coefficients.
The results exhibit a nice structure and will be presented in section
\ref{sec:infFF}.

\section{Infinite volume form factors}

\label{sec:infFF} First, let us comment on the identity operator.
As any multi-magnon diagonal matrix element of the identity operator
equals to $1$, matching it with the series (\ref{eq:Connected_expansion}),
one can easily derive that all infinite volume connected form factors
vanish except from the vacuum expectation value,
\begin{equation}
f^{\mathbb{I}}(\emptyset)=1\quad;\qquad f^{\mathbb{I}}(u_{1},...,u_{N})=0\quad,\quad N\geq1.
\end{equation}

We should also discuss separately the simple case of the vacuum expectation
values of spin chain operators. In the series (\ref{eq:Connected_expansion}),
the zero magnon diagonal matrix element only contains the vacuum expectation
value of the given operator in the infinite volume theory. So that,
for the length-$1$ and length-$2$ operators one can easily find
\begin{equation}
f^{o_{1}}(\emptyset)=f^{o_{2}^{1}}(\emptyset)=1\quad,\qquad f^{o_{2}^{2}}(\emptyset)=f^{o_{2}^{3}}(\emptyset)=f^{o_{2}^{4}}(\emptyset)=0.\label{eq:VEV_oi}
\end{equation}
It holds also in the general case. Let us take the operators $\rE_{n}^{ab}$,
($a,b=1,2$) as a basis on the local quantum space, and linearly extend
it to $l_{s}$ neighboring site. Then only one, among this $4^{l_{s}}$
basis element, has non-vanishing vacuum expectation value, namely
the one containing $\rE^{11}$ at each site.

In the rest of this section, we will perform the finite volume expansion
for the diagonal matrix elements of length-$1$ and length-$2$ operators.
We will discuss a simple example, namely the case of length-$1$ operator
with $2$ magnons in detail and present the results for more complicated
form factors.

\subsection{An example: length-$1$ operator with -$2$ magnons}

\label{sec:example} We consider the finite volume diagonal matrix
element for the operator $o_{1}(n)=\rE_{n}^{11}$ with two magnons
\begin{align}
\cF_{L}^{o_{1}}(u_{1},u_{2})=\frac{\langle u_{1},u_{2}|o_{1}(n)|u_{1},u_{2}\rangle}{\langle u_{1},u_{2}|u_{1},u_{2}\rangle}.
\end{align}
It has the following structure in finite volume
\begin{align}
\cF_{L}^{o_{1}}(u_{1},u_{2})=\frac{1}{\rho_{2}(\{1,2\})}\left(\rho_{2}(\{1,2\})+f^{o_{1}}(u_{2})\,\rho_{2}(\{1\})+f^{o_{1}}(u_{1})\,\rho_{2}(\{2\})+f^{o_{1}}(u_{1},u_{2})\right)\label{eq:o1Expd}
\end{align}
where $f^{o_{1}}(\mathbf{u})$ is to be identified with the connected
diagonal form factor of $o_{1}$ in the infinite volume theory (\ref{eq:connected_FF}).

We proceed as described in the previous sections. Using the solution
of QISP, we have
\begin{align}
\cF_{L}^{o_{1}}(u_{1},u_{2})=\frac{1}{t_{\mathbf{u}}(\theta_{n}^{+})}\frac{\langle u_{1},u_{2}|A(\theta_{n}^{+})|u_{1},u_{2}\rangle}{\langle u_{1},u_{2}|u_{1},u_{2}\rangle},\label{eq:FFo1}
\end{align}
where the denominator is the Gaudin norm (\ref{eq:norm}),
\begin{align}
\langle u_{1},u_{2}|u_{1},u_{2}\rangle=\left(\prod_{j=1}^{2}a(u_{j})d(u_{j})\right)\frac{1+(u_{1}-u_{2})^{2}}{(u_{1}-u_{2})^{2}}\,\rho_{2}(\{1,2\}).
\end{align}
From (\ref{eq:actAD}),
\begin{align}
\langle u_{1},u_{2}|A(\theta_{n}^{+})|u_{1},u_{2}\rangle= & \,\rM_{0}(\theta_{n}^{+})\langle u_{1},u_{2}|u_{1},u_{2}\rangle\\
+ & \,\rM_{1}(\theta_{n}^{+})\langle u_{1},u_{2}|u_{2},\theta_{n}^{+}\rangle+\rM_{2}(\theta_{n}^{+})\langle u_{1},u_{2}|u_{1},\theta_{n}^{+}\rangle.\nonumber
\end{align}
We introduce some notations in order to simplify the expressions.
Let us define
\begin{align}
\mathscr{C}_{\mathbf{u},\mathbf{v}}=\frac{\prod_{j=1}^{N}a(v_{j})d(u_{j})}{\det_{jk}\frac{1}{u_{j}-v_{k}+i}},
\end{align}
so that
\begin{align}
\langle\mathbf{v}|\mathbf{u}\rangle=\mathscr{C}_{\mathbf{u},\mathbf{v}}\,\det_{jk}\Omega(u_{j},v_{k}).
\end{align}
By perform the finite volume expansion for the three scalar products,
we obtain
\begin{align}
\langle u_{1},u_{2}|A(\theta_{n}^{+})|u_{1},u_{2}\rangle= & \,-\mathscr{C}_{\{u_{1},u_{2}\},\{u_{1},u_{2}\}}\rM_{0}(\theta_{n}^{+})\,\rho_{2}(\{1,2\})\label{eq:o1calc}\\
- & \, i\mathscr{C}_{\{u_{1},u_{2}\},\{u_{1},\theta_{n}^{+}\}}\rM_{2}(\theta_{n}^{+})\Omega(u_{2},\theta_{n}^{+})\,\rho_{2}(\{1\})\nonumber \\
- & \, i\mathscr{C}_{\{u_{1},u_{2}\},\{u_{2},\theta_{n}^{+}\}}\rM_{1}(\theta_{n}^{+})\Omega(u_{1},\theta_{n}^{+})\,\rho_{2}(\{2\})\nonumber \\
- & \,\phi_{12}\left(\mathscr{C}_{\{u_{1},u_{2}\},\{u_{2},\theta_{n}^{+}\}}\Omega(u_{2},\theta_{n}^{+})\rM_{1}(\theta_{n}^{+})+\mathscr{C}_{\{u_{1},u_{2}\},\{u_{1},\theta_{n}^{+}\}}\Omega(u_{1},\theta_{n}^{+})\rM_{2}(\theta_{n}^{+})\right)\nonumber
\end{align}
Plugging (\ref{eq:o1calc}) into (\ref{eq:FFo1}) and comparing to the
expansion (\ref{eq:o1Expd}), we obtain the expression for the various
form factors in infinite volume
\begin{align}
f^{o_{1}}(u_{1})= & \, i\frac{\mathscr{C}_{\{u_{1},u_{2}\},\{u_{2},\theta_{n}^{+}\}}}{\mathscr{C}_{\{u_{1},u_{2}\},\{u_{1},u_{2}\}}}\,\frac{\rM_{1}(\theta_{n}^{+})}{\rM_{0}(\theta_{n}^{+})}\,\Omega(u_{1},\theta_{n}^{+})\label{eq:fo1}\\
f^{o_{1}}(u_{2})= & \, i\frac{\mathscr{C}_{\{u_{1},u_{2}\},\{u_{1},\theta_{n}^{+}\}}}{\mathscr{C}_{\{u_{1},u_{2}\},\{u_{1},u_{2}\}}}\,\frac{\rM_{2}(\theta_{n}^{+})}{\rM_{0}(\theta_{n}^{+})}\,\Omega(u_{2},\theta_{n}^{+})\nonumber \\
f^{o_{1}}(u_{1},u_{2})= & \,\frac{\phi_{12}}{\mathscr{C}_{\{u_{1},u_{2}\},\{u_{1},u_{2}\}}\,\rM_{0}(\theta_{n}^{+})}\left(\mathscr{C}_{\{u_{1},u_{2}\},\{u_{2},\theta_{n}^{+}\}}\Omega(u_{2},\theta_{n}^{+})\rM_{1}(\theta_{n}^{+})\right.\nonumber \\
 & \,\left.+\mathscr{C}_{\{u_{1},u_{2}\},\{u_{1},\theta_{n}^{+}\}}\Omega(u_{1},\theta_{n}^{+})\rM_{2}(\theta_{n}^{+})\right)\nonumber
\end{align}
Substituting the explicit expressions in (\ref{eq:fo1}) and, at the
end, taking the homogeneous limit $\theta_{n}\to0$, we obtain very
compact results for the infinite volume connected form factors,
\begin{align}
f^{o_{1}}(u_{k})= & \,\frac{1}{u_{k}^{2}+1/4},\qquad k=1,2\\
f^{o_{1}}(u_{1},u_{2})= & \,\left(\frac{1}{u_{1}^{2}+1/4}+\frac{1}{u_{2}^{2}+1/4}\right)\frac{2}{1+(u_{1}-u_{2})^{2}}.\nonumber
\end{align}

\subsection{Length-$1$ operator with $N$ magnons}

\begin{figure}[h!]
\begin{centering}
\includegraphics[scale=0.5]{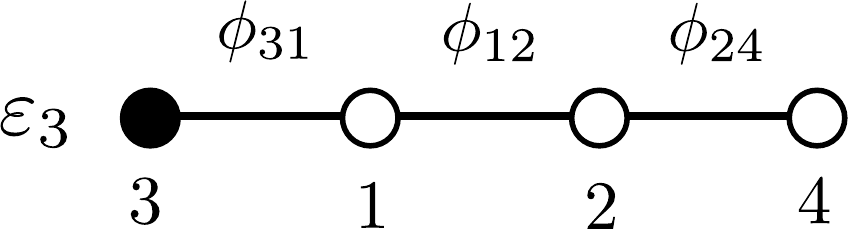} \caption{Diagrammatic representation of one term in (\ref{eq:o1N}) with $4$
magnons.}
\label{fig:fo1}
\par\end{centering}
\centering{}
\end{figure}

We can perform the same calculation as in the previous subsection
and extract the form factors with more magnons. The process becomes
cumbersome for higher number of particles. However, from the first
few magnon cases, we are able to observe a nice pattern of the connected
form factors. The $N$-magnon connected diagonal form factor for $o_{1}(n)$
is given as
\begin{align}
f^{o_{1}}(u_{1},...,u_{N})=\varepsilon_{1}\,\phi_{12}\,\phi_{23}...\phi_{N-1,N}+\text{permutations}\label{eq:o1N}
\end{align}
where $\varepsilon_{k}$ is the energy of the magnon with rapidity
$u_{k}$ and $\phi_{jk}$ can be seen as some ``propagator'' defined
as
\begin{align}
\varepsilon_{k}=\varepsilon(u_{k})=\frac{1}{u_{k}^{2}+1/4},\qquad\phi_{jk}=\frac{2}{1+(u_{j}-u_{k})^{2}}\quad,\quad j\neq k.
\end{align}
The expression (\ref{eq:o1N}) can be represented by the diagrams
in Fig.(\ref{fig:fo1}). Each node is labeled by a number from $1$
to $N$. The leftmost node is associated with the energy of its label.
The lines between two neighboring nodes are associated with a propagator.
Multiplying the factors we obtain the value of the diagram. Summing
over all the permutations of the labeling gives the result for infinite
volume form factor $f^{o_{1}}$. The result for an $N$ magnon state
is thus a sum over $N!$ terms.

The structure of $f^{o_{1}}$ is exactly the structure of the connected
form factors of conserved charge densities (\ref{eq:FF_conn_conservedQ}).
This is not surprising, since $o_{1}(n)=\frac{1}{2}(\mathbb{I}+\sigma_{n}^{z})$
is indeed a length-$1$ conserved charge density of the Heisenberg
spin chain. The nice feature is that once we know the one particle
eigenvalue $q(u)$ of the charge, we can immediately write down the
expression for the corresponding infinite volume form factors. We
remark here that our result (\ref{eq:o1N}) is consistent with the
determinant formula of \cite{Kitanie:FF}.

\subsection{Length-$2$ form factors of $N$ magnons}
\begin{figure}[h!]
\begin{centering}
\includegraphics[scale=0.5]{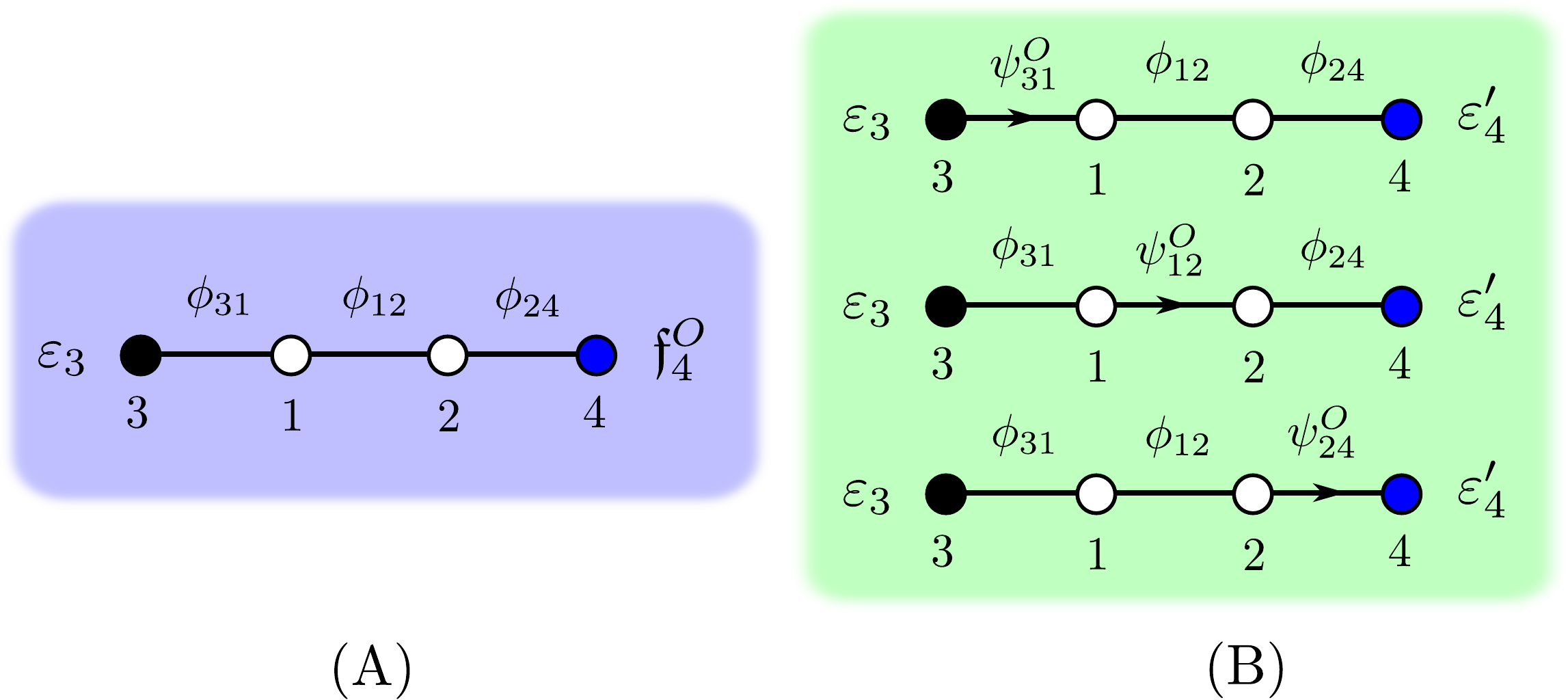} \caption{Different kinds of diagrams encoding length-$2$ form factors.}
\label{fig:fo2}
\par\end{centering}
\end{figure}

The calculation of infinite volume matrix elements can be performed
following the same line as in section \ref{sec:example} but the process
is more involved.
Nevertheless, we again found some patterns for the various matrix elements
which we present below. The structure for the length-$2$ operators
can be encoded into diagrams similar to the one in Fig.(\ref{fig:fo1}).
However, in this case we have two types of them, as are shown in Fig.(\ref{fig:fo2}).

Each diagonal matrix element is given by two kinds of diagrams. This structure was tested for all length-2 operators up to 5 magnons and we conjecture that it holds for any number of magnons. The
first kind is depicted in the blue region. We label the nodes by number
from $1$ to $N$. The leftmost node is associated with $\varepsilon(u_{j})$
while the rightmost node is associated with a function denoted by
$\mathfrak{f}^{O}(u_{k})$, and it depends on the operator. The other
kind, depicted in the green region, is more interesting. The leftmost
and rightmost nodes are associated with $\varepsilon_{j}$ and $\varepsilon'_{k}$
where $\varepsilon'(u)=\frac{\partial}{\partial u}\varepsilon(u)$.
In addition, for a given label of the nodes, one needs to sum over
the diagrams which replaces one of the propagators by a ``directed
propagator'', $\psi_{ij}^{O}=\psi^{O}(u_{i},u_{j})$. The directed
propagator is antisymmetric with respect to its arguments $\psi_{ij}^{O}=-\psi_{ji}^{O}$
and its explicit form depends on the operator under consideration.

In summarizing, the infinite volume diagonal matrix element of a length-2
operator, $f^{O}(u_{1},\cdots,u_{N})$, is characterized by two functions
$\mathfrak{f}^{O}(u)$ and $\psi^{O}(u,v)$. The result for $N$-magnon
is given by
\begin{align}
f^{O}(u_{1},...,u_{N})= & \,\left(\varepsilon_{1}\,\phi_{12}...\phi_{N-1,N}\,\mathfrak{f}_{N}^{O}+\text{permutations}\right)\nonumber \\
+ & \,\left(\sum_{i=1}^{N-1}\varepsilon_{1}\,\phi_{12}...\psi_{i,i+1}^{O}...\phi_{N-1,N}\,\varepsilon'_{N}+\text{permutations}\right)\label{eq:Length2_IVFF}
\end{align}
We list the data for $o_{2}^{i}$ (\ref{eq:o1},\ref{eq:o2}) is the
following:
\begin{align}
&\mathfrak{f}^{o_{2}^{1}}(u)=2\qquad & &\psi^{o_{2}^{1}}(u,v)=-(u-v)(uv-1/4)\phi(u,v)&\nonumber \\
&\mathfrak{f}^{o_{2}^{2}}(u)=-\frac{u-i/2}{u+i/2}\qquad& & \psi^{o_{2}^{2}}(u,v)=(u-v)(u-i/2)(v-i/2)\phi(u,v)\nonumber& \\
&\mathfrak{f}^{o_{2}^{3}}(u)=-\frac{u+i/2}{u-i/2}\qquad& & \psi^{o_{2}^{3}}(u,v)=(u-v)(u+i/2)(v+i/2)\phi(u,v)\nonumber& \\
&\mathfrak{f}^{o_{2}^{4}}(u)=0\qquad& & \psi^{o_{2}^{4}}(u,v)=-(u-v)(uv-1/4)\phi(u,v)\label{eq:dataf}&
\end{align}
Let us comment on this results (\ref{eq:dataf}). These data for the
operators can be read off simply from the computation of $2$ magnon
case. Therefore, one should simply compute the $2$-magnon matrix
elements and perform the finite volume expansion to extract the data.
Once the functions in (\ref{eq:dataf}) are known, we can write down
any diagonal form factor of length-$2$ operators in the infinite
volume. Any length-$2$ operator is a linear combination of the identity
operator, $o_{1}$ and $o_{2}^{i}$,
\begin{align}
\mathcal{O}_n=b\,\mathbb{I}+c_{0}\,o_{1}(n)+\tilde{c}_0\,o_1(n+1)+\sum_{i=1}^{3}c_{i}\, o_{2}^{i}(n)
\end{align}
where $b$, $c_{0}$ and $c_{i}$ $(i=1,2,3)$ are
some numbers. Then the data of $\mathcal{O}$ is simply given by
\begin{align}
\mathfrak{f}^{\mathcal{O}}(u)=c_{0}+\tilde{c}_0+\sum_{i=1}^{3}c_{i}\,\mathfrak{f}^{o_{2}^{i}}(u),\qquad\psi^{\mathcal{O}}(u,v)=\sum_{i=1}^{3}c_{i}\,\psi^{o_{2}^{i}}(u,v),\label{eq:dataO}
\end{align}
and its vacuum expectation value is
\begin{equation}
f^{\mathcal{O}}(\emptyset)=b+c_{0}+\tilde{c}_0+c_{1}.\label{eq:VEV_length2}
\end{equation}
For example, the operator $o_{2}^{4}$ is not independent
\[
o_{2}^{4}(n)=\mathbb{I}-o_{1}(n)-o_{1}(n+1)+o_{2}^{1}(n).
\]
Note that by translational invariance, $o_1(n)$ gives the same result as $o_1(n+1)$ when computing the form factors. It is easy to check that this resolution is consistent with (\ref{eq:VEV_oi},\ref{eq:dataf}).
The diagonal matrix elements of the operators $o_{2}^{2}$ and $o_{2}^{3}$
are related by complex conjugation which is also manifest in (\ref{eq:dataf}).

\subsection{Examples of length-$2$ operators}

We compute two examples below. First, the matrix elements of the length-$2$
conserve charge density, which is the permutation operator $\rP_{k,k+1}$,
or equivalently the Hamiltonian density $\rH_{k,k+1}=\rI_{k,k+1}-\rP_{k,k+1}$.
We will see that the data for the permutation operator simplifies
and the final result takes exactly the form predicted in (\ref{eq:FF_conn_conservedQ}).
This is a non-trivial check of our functions (\ref{eq:dataf}). Another
example is an HHL three-point function, with the light operator being
the rotated BMN vacuum of length $4$.

\paragraph{Permutation operator}

The permutation operator $\rP_{k,k+1}$ is a length-$2$ operator
of the Heisenberg spin chain. It can be written in terms of the operators
$o_{2}^{i}$ with equal weights
\begin{align}
\rP_{k,k+1}=\sum_{i,j=1}^{2}\rE_{k}^{ij}\,\rE_{k+1}^{ji}=\sum_{i=1}^{4}o_{2}^{i}.
\end{align}
According to (\ref{eq:dataO}), the data of permutation operator is
given by
\begin{align}
\mathfrak{f}^{\rP}(u)=\varepsilon(u)=\frac{1}{u^{2}+1/4},\qquad\psi^{\rP}(u,v)=0, \qquad f^P(\emptyset)=1.\label{eq:dataP2}
\end{align}
The infinite volume form factor (\ref{eq:Length2_IVFF}) with the
entires (\ref{eq:dataP2}) has the structure as a conserved charge
should have (\ref{eq:FF_conn_conservedQ}), with the one particle eigenvalue
of the corresponding charge being $-\varepsilon(u)$.

\paragraph{A three-point function with $\mathcal{O}_{\alpha}=\Tr\tilde{Z}\tilde{Z}\tilde{Z}\tilde{Z}$}

As another example, we compute a HHL three-point function with the
light operator being $\mathcal{O}_{\alpha}=\tilde{\mathcal{O}}=\Tr\tilde{Z}\tilde{Z}\tilde{Z}\tilde{Z}$,
which is the rotated BMN vacuum. The scalar field $\tilde{Z}$ is
defined as
\begin{align}
\tilde{Z}=Z+\bar{Z}+i(X+\bar{X}).
\end{align}
Following our strategy, we first map the field theory operator to
the spin chain operator $\tilde{\mathcal{O}}$ using (\ref{mapping}). Then
we write the operator $\tilde{\mathcal{O}}$ in terms of linear combinations
of the basis operators $o_{1},o_{2}^{i}$ ($i=1,...,3$). This enable
us to write down the data for the operator $\tilde{\mathcal{O}}$ and thus
the infinite volume form factor. We have\footnote{Factor of 4 appeared due to the symmetry of the $\tilde{\mathcal{O}}$}
\beq
{\cal \tilde{\mathcal{O}}}_n = 4( \mathbb{I} -2o_1(n)-2o_1(n+1)+4o_2^1(n)-o_2^2(n)-o_2^3(n)).
\label{Ztilde}
\eeq
Using the representation \ur{Ztilde} and also the data \ur{eq:dataf},
we get for the operator $\tilde{\mathcal{O}}$
\beq
\mathfrak{f}^{\tilde{\mathcal{O}}}(u)=\frac{96u^2+8}{4u^2+1},\qquad
\psi^{\tilde{\mathcal{O}}}(u,v)=-6(u-v)(4uv-1)\phi(u,v).
\eeq
The corresponding infinite volume form factors can be obtained from the general prescription \ur{eq:Length2_IVFF}
for the length-2 form factors.

\section{Matrix elements at one loop}

\label{sec:loopFF} In this section, we generalize the above considerations
from tree level to one loop. We show that the form factors at
one loop can again be written in terms of a finite number of ``building
blocks''. These building blocks are matrix elements of the inhomogeneous
Heisenberg XXX$_{1/2}$ spin chain with the inhomogeneities fixed
to some specific values, called the BDS values, and can be written in terms of scalar products \ur{eq:mGN}
for which one can perform the finite volume expansion, as at
the tree level.

There are several new features for three-point functions at higher
loops. The dilatation operator in the $\mathfrak{su}(2)$ sector is
no longer the Hamiltonian of the Heisenberg XXX$_{1/2}$ spin chain,
but becomes long-range interacting, called the BDS spin chain \cite{Beisert:2004hm}.
Therefore the two large operators correspond to the eigenvectors of
the BDS spin chain. The BDS spin chain is related to a special inhomogeneous
Heisenberg XXX$_{1/2}$ spin chain by a unitary transformation \cite{Bargheer:2009xy,Jiang:OneLoop}.
Therefore its eigenstates can be obtained from the eigenstates of
the inhomogeneous Heisenberg spin chain by performing the unitary
transformation, namely $|\mathbf{u}\rangle_{\text{BDS}}=\rS|\mathbf{u};\bm{\theta}^{\text{BDS}}\rangle$.
The BDS values of the inhomogeneities are given by
\begin{align}
\theta_{k}^{\text{BDS}}=2g\sin\frac{2\pi k}{L},\quad k=1,\cdots,L
\end{align}
where $g$ is the coupling constant. The unitary operator $\rS$ has
been worked out in \cite{Jiang:OneLoop} up to $g^{2}$ order
\begin{align}
\rS=\exp i\sum_{k=1}^{L}\left[\nu_{k}\rH_{k}+\frac{i}{2}\rho_{k}[\rH]_{k}\right],\label{fixing}
\end{align}
where $\rH_{k}\equiv\mathrm{I}_{k,k+1}-\rP_{k,k+1}$ and $[\rH]_{k}\equiv[\rH_{k},\rH_{k+1}]$.
The parameters $\nu_{k}$ and $\rho_{k}$ are related to the inhomogeneities
by
\begin{align}
\nu_{k}=-\sum_{j=1}^{k}\theta_{j},\quad\rho_{k}=2g^{2}k-\theta_{k}\nu_{k}-\sum_{j=1}^{k}\theta_{j}^{2},\quad k=1,\cdots,L.
\end{align}
It is obvious that $\nu_{k}\sim g$ and $\rho_{k}\sim g^{2}$.

\begin{figure}[h!]
\begin{centering}
\includegraphics[scale=0.5]{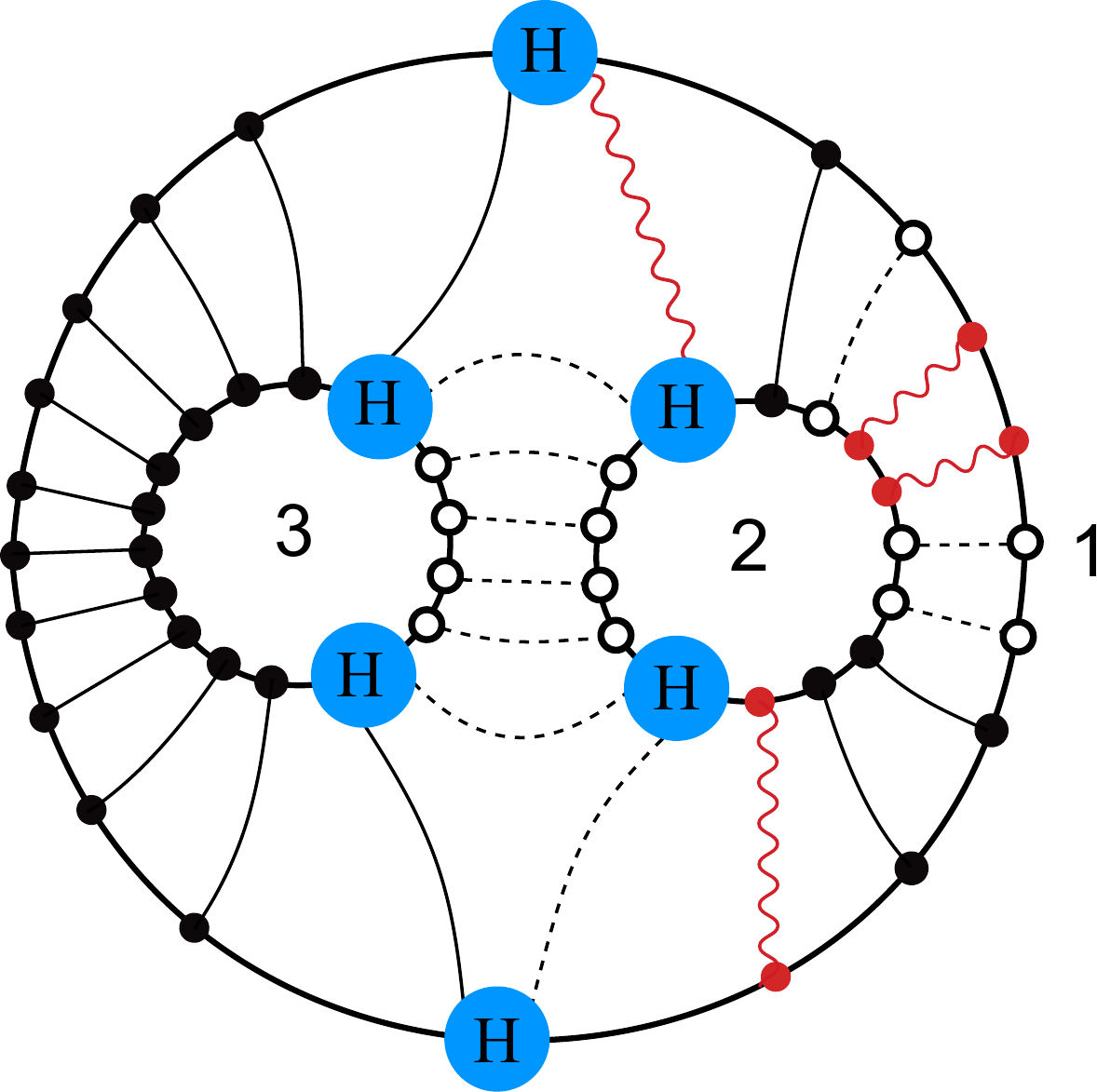} \caption{The quantum corrections are taken into account by operator insertions
at the splitting points. At one-loop in the $\mathfrak{so}(6)$ sector,
the insertion takes the form of one-loop Hamiltonian density.}
\label{loop}
\par\end{centering}
\centering{}
\end{figure}

Another new feature is that the quantum corrections manifest themselves
as operator insertions at the splitting points \cite{Okuyama-Tseng,Alday:3pt},
as is shown in Fig.\ref{loop}. The operator insertions take the form of the Hamiltonian density of the spin chain.
In our set up we need insertions for the $\mathfrak{so}(6)$ sector, since
the light operator belongs to $\mathfrak{so}(4)$, which is not closed at one loop.

At one loop level, we need to take into account the aforementioned
new features. In what follows, we first discuss the effect of the
operator insertions and show that the three-point function under consideration
can be reduced into the calculation of correlation functions of the
BDS spin chain. Then we consider the effect of the $\rS$ operator
on the spin operators and reduce the correlation functions of BDS
spin chain into the correlation functions of the inhomogeneous Heisenberg
XXX$_{1/2}$ spin chain. As we discussed in the tree level case, the
matrix elements of the inhomogeneous Heisenberg spin chain can be
written in terms of scalar products (\ref{eq:mGN}), and we
can perform the finite volume expansion.

\subsection{The effects of one-loop operator insertions}

For each spin chain state, there are two operator insertions at the
two splitting points. We first discuss the effects of insertions for
the ``light'' operator $\mathcal{O}_{\alpha}$. The one-loop insertion
takes the form of the Hamiltonian density
\begin{align}
\rH_{l}^{\text{so(6)}}=\rK_{l,l+1}+2\rI_{l,l+1}-2\rP_{l,l+1}
\end{align}
where $\rI_{l,l+1}$, $\rP_{l,l+1}$ and $\rK_{l,l+1}$ are the identity,
permutation and trace operators. They act on the $\mathfrak{so}(6)$
spin chain states as
\begin{align}
 & \rI_{l,l+1}|...\phi_{l}^{i}\phi_{l+1}^{j}...\rangle=|...\phi_{l}^{i}\phi_{l+1}^{j}...\rangle,\label{IPK}\\
 & \rP_{l,l+1}|...\phi_{l}^{i}\phi_{l+1}^{j}...\rangle=|...\phi_{l}^{j}\phi_{l+1}^{i}...\rangle,\nonumber \\
 & \rK_{l,l+1}|...\phi_{l}^{i}\phi_{l+1}^{j}...\rangle=\delta_{ij}\sum_{k=1}^{6}|...\phi_{l}^{k}\phi_{l+1}^{k}...\rangle.\nonumber
\end{align}

\begin{figure}[h!]
\begin{centering}
\includegraphics[scale=0.5]{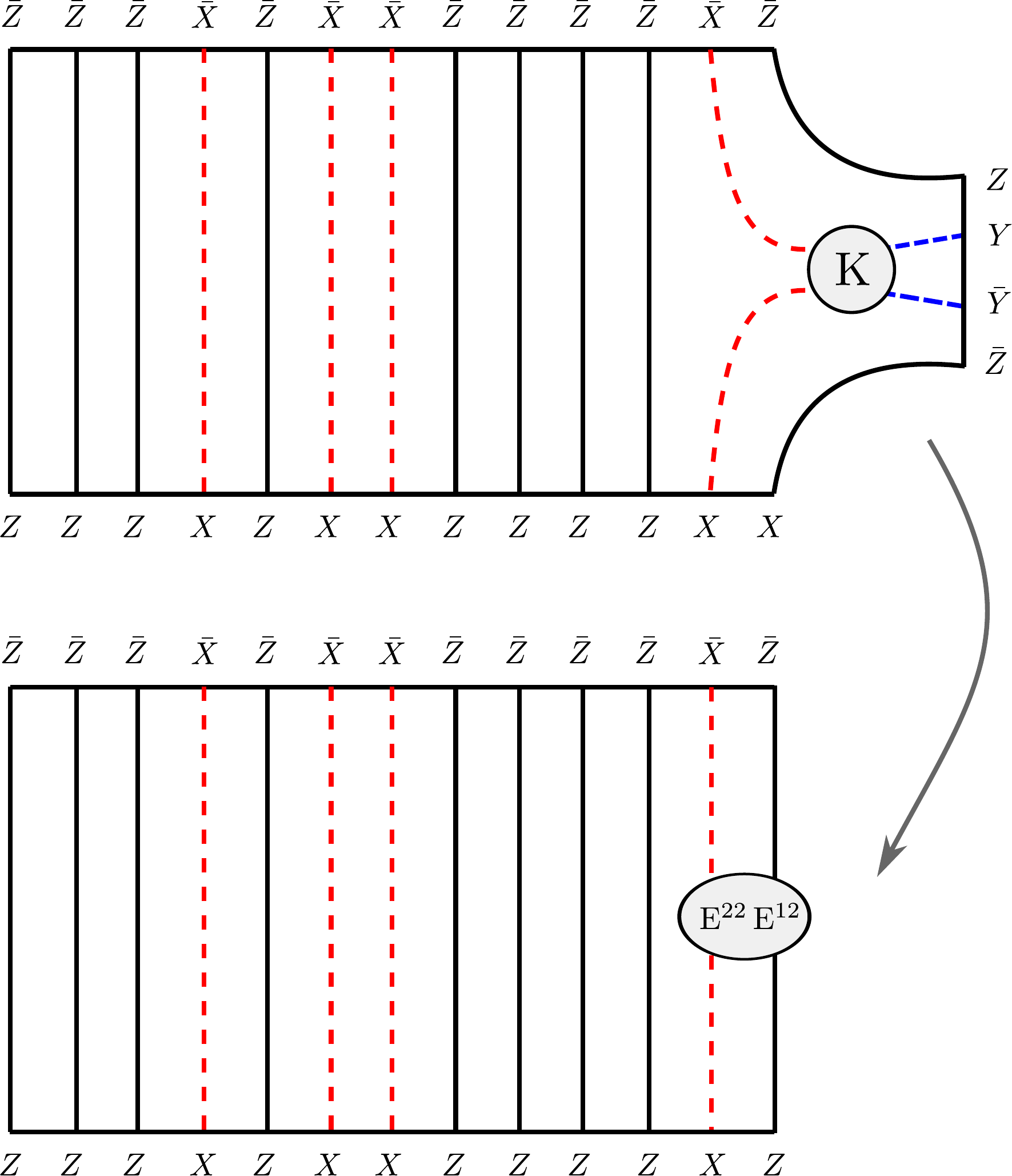} \caption{An example of mapping the light operator to the spin chain operator at one-loop in the presence of one-loop insertion.}
\label{YY}
\par\end{centering}
\centering{}
\end{figure}

At one loop level, the light operator should be an eigenvalue of the two-loop
dilatation operator. The $\mathfrak{so}(6)$
sector is closed only at one-loop so in principle one needs fields
outside the $\mathfrak{so}(6)$ sector, like fermionic fields, to
construct the eigenstates of the two loop dilatation operator. However,
when computing the three-point functions, the Feymann diagrams, involving fields apart from $\{X,Z,Y,\bar{X},\bar{Z},\bar{Y}\}$,
will not contribute at one loop order. Thus, the only new fields which we have to take into account at one loop are $Y$ and $\bar Y$.
It's easy to see from Fig.\,\ref{YY} that the light operator, in the presence of one-loop insertions, can again be written in terms of local spin chain operators $\sigma^{\pm},\sigma^z$, due to the fact that the heavy operators are still can be expressed as $\mathfrak{su}(2)$ spin chain states. Therefore, our considerations before can be generalized here.
To summarize, the one-loop structure
constants can be recast to the calculation of matrix elements
of the BDS spin chain $\phantom{|}_{\text{BDS}}\langle\mathbf{u}|\hat{O}(\sigma^{\pm},\sigma^{z};g^{2})|\mathbf{u}\rangle_{\text{BDS}}$.

\begin{figure}[h!]
\begin{centering}
\includegraphics[scale=0.5]{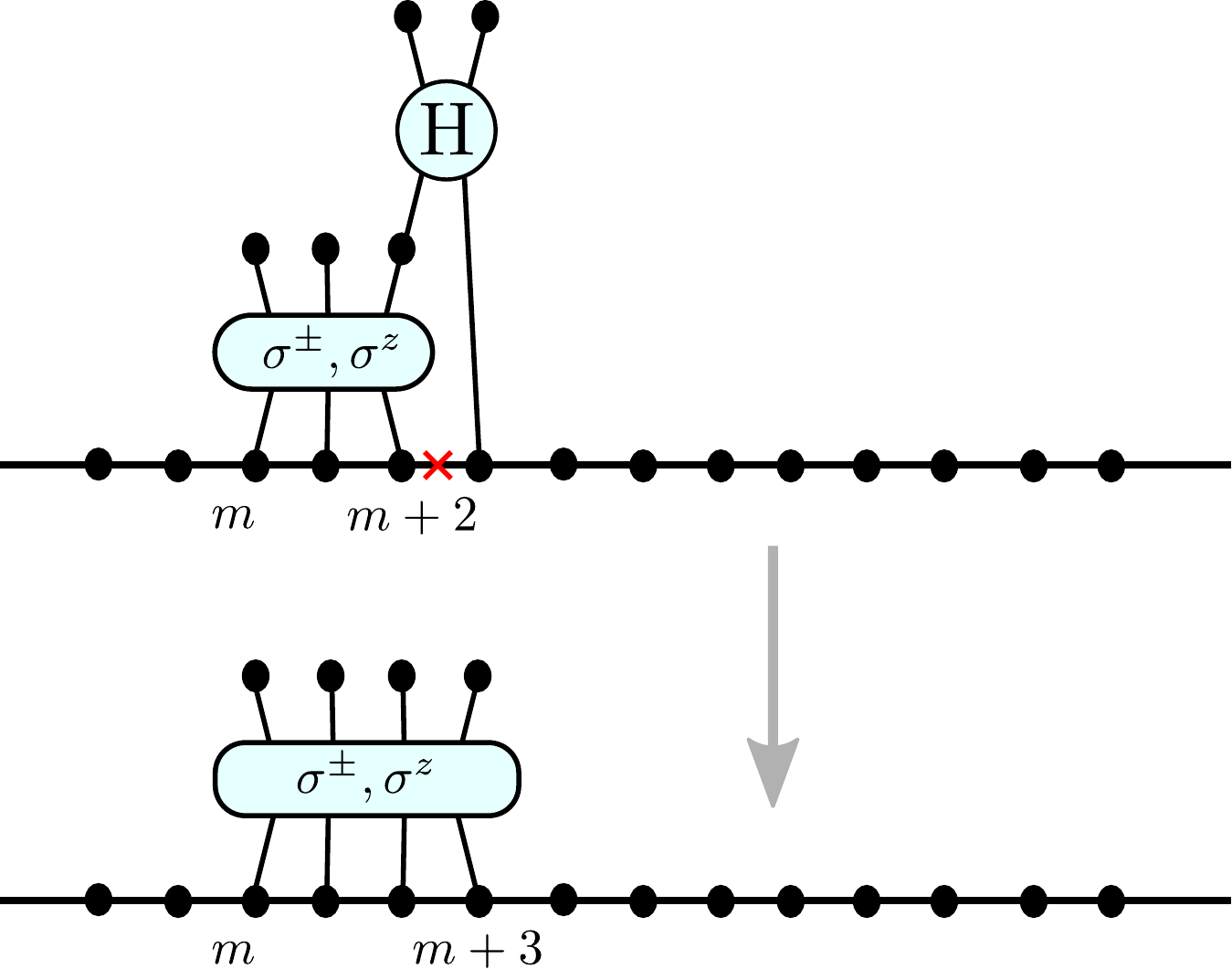} \caption{The effect of operator insertions for the heavy operators. They increase
the length of the spin operator by $1$. The red cross denotes the
splitting point.}
\label{increase}
\par\end{centering}
\centering{}
\end{figure}

Now we consider the operator insertions for the heavy states. The
effect of these insertions is increasing the length of the spin chain
operator, as is shown in (\ref{increase}).

This can be seen easily by noticing that
\begin{align}
\rP_{k,k+1}=\sum_{i,j=1}^{2}\rE_{m}^{ij}\otimes\rE_{m+1}^{ji}
\end{align}
For example, we have the following
\begin{align}
\star\,\rE_{m}^{11}\rH_{m}\,\star=(\star\,\rE_{m}^{11}\,\star)-(\star\,\rE_{m}^{11}\rE_{m+1}^{11}\,\star)-(\star\,\rE_{m}^{11}\rE_{m+1}^{12}\,\star)
\end{align}
where the star stands for some strings of operators.

\subsection{The effects of the unitary $\rS$ operator}

\begin{figure}[h!]
\begin{centering}
\includegraphics[scale=0.5]{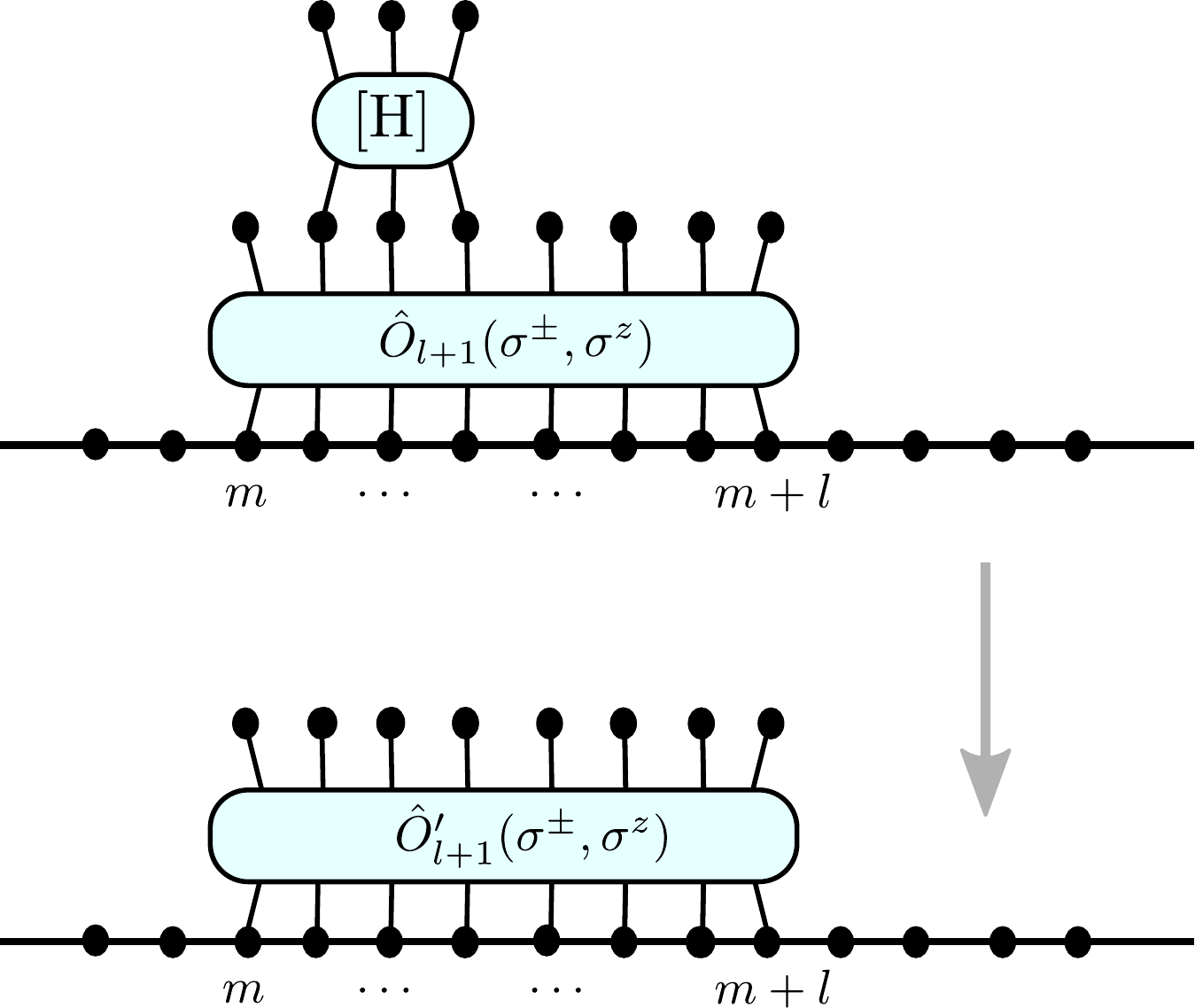} \caption{The length preserving action generated by $[\rH]_{k}$ on the spin
chain vertex.}
\label{bulkH}
\par\end{centering}
\centering{}
\end{figure}

In this subsection, we discuss the action of unitary operator $\rS$
on the spin operators. We are interested in the following quantity
\begin{align}
\langle\mathbf{u};\bm{\theta}^{\text{BDS}}|\rS^{-1}\hat{O}_{l+1}(\sigma^{\pm},\sigma^{z})\rS|\mathbf{u};\bm{\theta}^{\text{BDS}}\rangle.
\end{align}
The $\rS$ operator takes an exponential form $\rS=\exp\hat{\rF}$,
thus we have
\begin{align}
\rS^{-1}\hat{O}_{l+1}(\sigma^{\pm},\sigma^{z})\rS=\hat{O}_{l+1}(\sigma^{\pm},\sigma^{z})-[\hat{\rF},\hat{O}_{l+1}(\sigma^{\pm},\sigma^{z})]+\frac{1}{2}[\hat{\rF},[\hat{\rF},\hat{O}_{l+1}(\sigma^{\pm},\sigma^{z})]]+\mathcal{O}(g^{3})
\end{align}
where we have truncated up to $\mathcal{O}(g^{2})$ order. The action
of $\rS$ operator on the spin chain operator can be divided into
two types. The first type is length preserving, it originates from
the operators $\rH_{k}$ and $[\rH]_{k}$ that act within the range
of the spin chain operator $\hat{O}_{l+1}$, which gives rise to an
operator with the same length, this is depicted in Fig.\,\ref{bulkH}.

\begin{figure}[h!]
\begin{centering}
\includegraphics[scale=0.5]{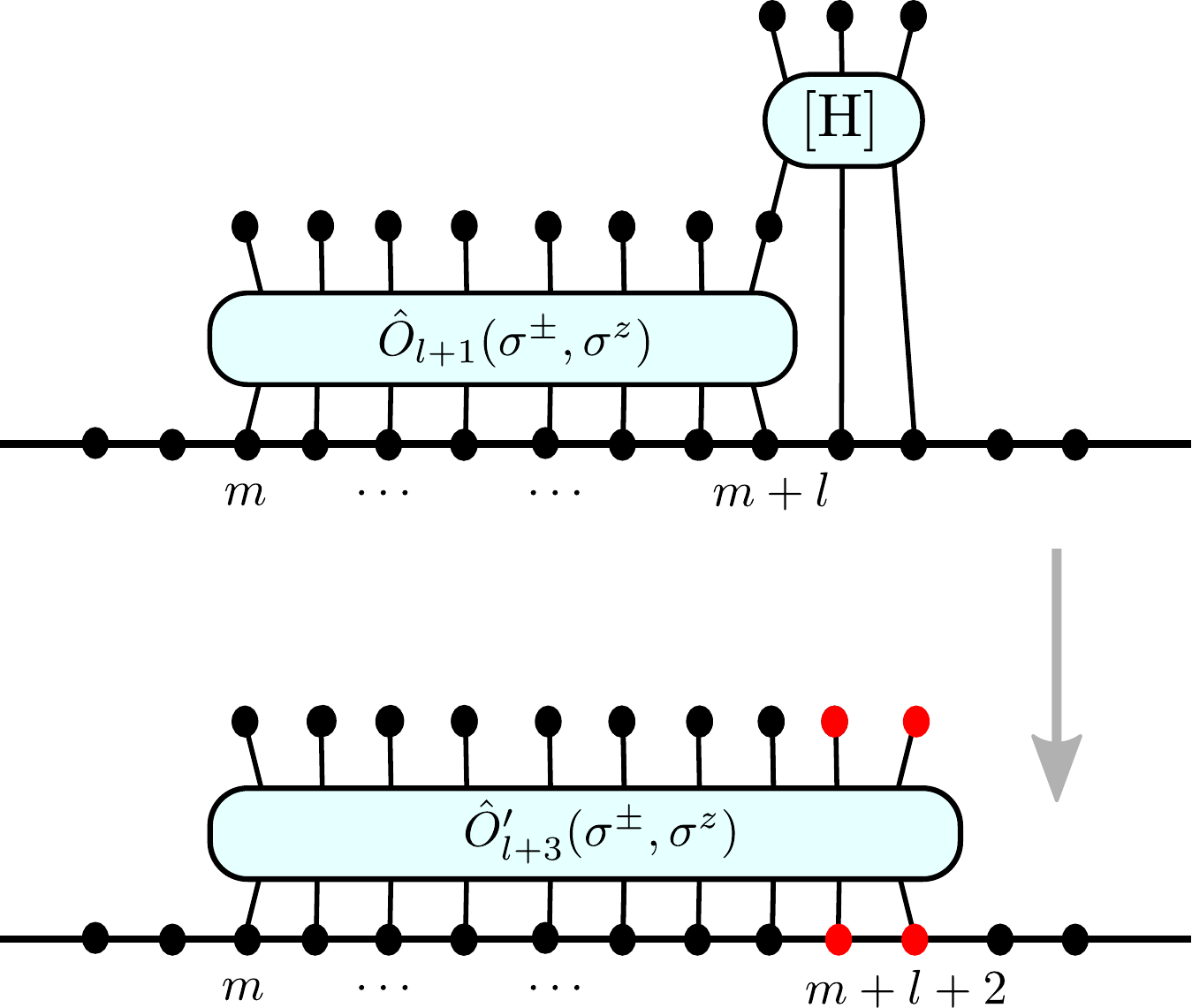} \caption{The length changing action generated by $[\rH]_{k}$ on the spin chain
vertex. In this example, it increases the length of the spin chain
operator by $2$.}
\label{boundaryH1}
\par\end{centering}
\centering{}
\end{figure}

The other type of the action increases the length of the operator
by $1$ or $2$, which are generated from the operators at the boundary
of the spin chain operator. There are two kinds of length changing
processes at one loop. One process is generated by a single $\rH_{k}$
or $[\rH]_{k}$, which is given in Fig.\,\ref{boundaryH1}.

\begin{figure}[h!]
\begin{centering}
\includegraphics[scale=0.5]{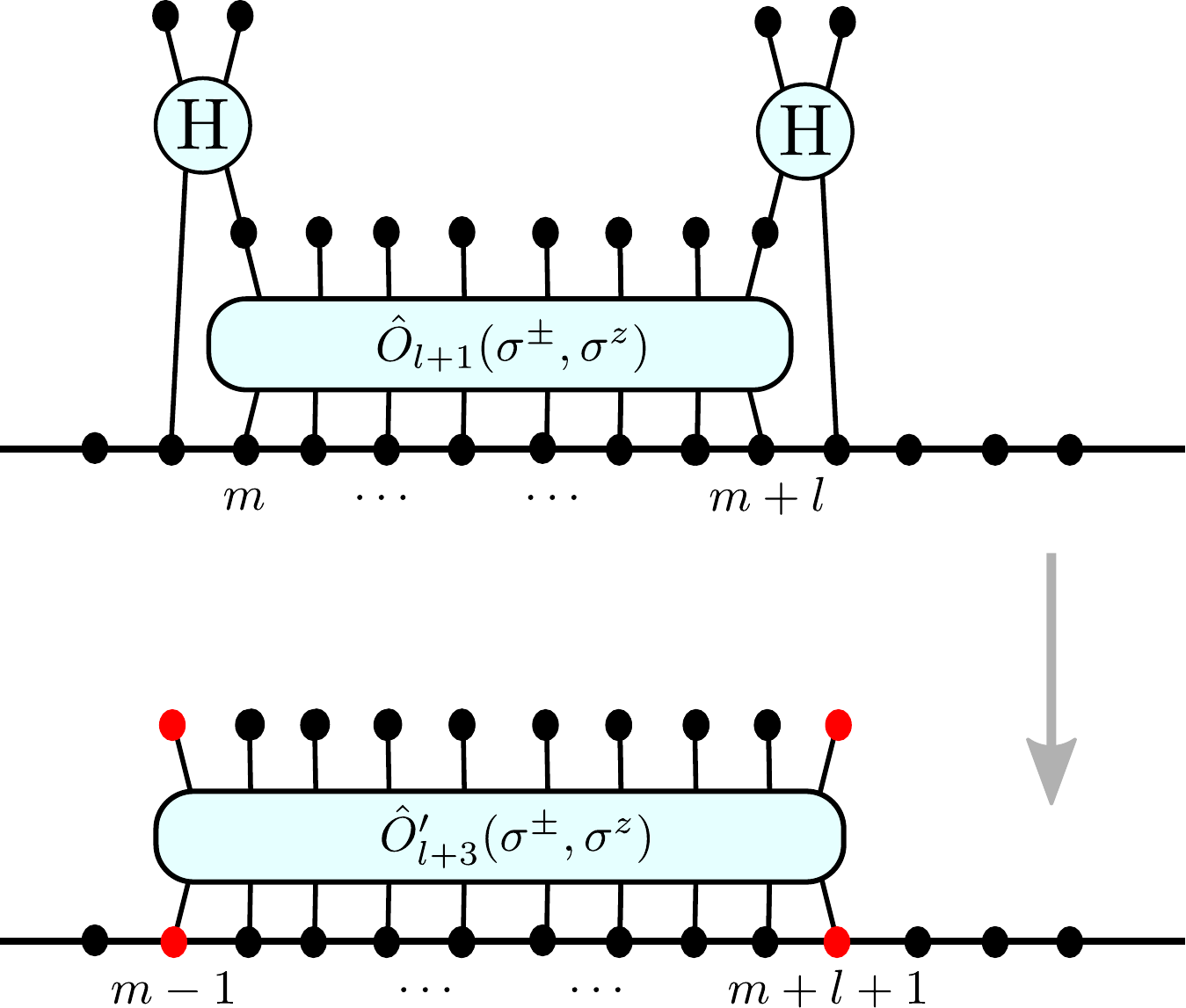} \caption{The length changing action generated by two $\rH_{k}$'s on both ends
of the spin chain vertex. The length of the spin operator also increases
by $2$ in this example.}
\label{boundaryH2}
\par\end{centering}
\centering{}
\end{figure}

The other process is generated by two $\rH_{k}$'s, one example of
which is given in Fig.\,\ref{boundaryH2}.

From our analysis we see that the action of the $\rS$ operator on
the spin chain operators, in general, increases the length of the
spin chain operator. Up to $\mathcal{O}(g^{2})$ order, the length
of the operator increases at most by $2$.
\begin{align}
\rS^{-1}\,\hat{O}_{l}(\sigma^{\pm},\sigma^{z})\,\rS=\hat{O}'_{l}(\sigma^{\pm},\sigma^{z})+\hat{O}'_{l+1}(\sigma^{\pm},\sigma^{z})+\hat{O}'_{l+2}(\sigma^{\pm},\sigma^{z})+\mathcal{O}(g^{3})
\end{align}
This implies that in order to compute the form factor of length $l$
operator for BDS spin chain, we need to compute the form factors of
length $l+2$, $l+1$ and $l$ operators in the inhomogeneous Heisenberg
XXX$_{1/2}$ spin chain.

Once we write the three-point function in terms of matrix elements
of the inhomogeneous XXX$_{1/2}$ spin chain, we can perform the finite
volume expansion and organize the results in the form conjectured
in \cite{Bajnok:2014sza}. At one loop, the matrix element of Gaudin
norm is modified. The equations (\ref{eq:Phik}), (\ref{eq:pu}) and
(\ref{eq:rhon}) are still valid, but the eigenvalues $a(u)$ and
$d(u)$ are corrected
\begin{align}
a(u)= & \,\prod_{k=1}^{L}(u-\theta_{k}^{\text{BDS}}+i/2)=x(u+i/2)^{L}+\mathcal{O}(g^{2L}),\\
d(u)= & \,\prod_{k=1}^{L}(u-\theta_{k}^{\text{BDS}}+i/2)=x(u-i/2)^{L}+\mathcal{O}(g^{2L}),\nonumber
\end{align}
where $x(u)$ is the Zhukowsky map given by
\begin{align}
x(u)+\frac{g^{2}}{x(u)}=u.
\end{align}
By replacing
\begin{align}
p(u_{k})=\frac{u_{k}+i/2}{u_{k}-i/2}\longrightarrow\,\frac{x(u_{k}+i/2)}{x(u_{k}-i/2)}\label{eq:puloop}
\end{align}
in (\ref{eq:Phik}) and expanding the result up to $\mathcal{O}(g^{2})$
order, we obtain the Gaudin norm at one-loop, $\rho_{n}^{\text{1-loop}}$.
In fact, the replacement (\ref{eq:puloop}) gives the correct Gaudin
norm up to wrapping orders\cite{GV,Jiang:OneLoop}. Our conclusion
is that the structure conjectured in \cite{Bajnok:2014sza} is also
valid at one loop level with respect to the one-loop Gaudin norm.

Of course the coefficients or the infinite volume form factors at
one-loop will be more complicated. It is an interesting question to
see how the infinite volume form factors are deformed at one loop
and whether it is possible to bootstrap to all loops.

\section{Conclusion}

In this paper we considered symmetric HHL correlators at weak coupling of the $\mathfrak{su}(2)$ sector of ${\cal N}=4$ SYM theory.
Using the formalism of computing matrix element of XXX${}_{1/2}$ spin chain (see e.g. \cite{Kitanie:FF}),
we showed that at tree level, as well as at one loop, the finite volume dependence (up to the wrapping corrections) of this kind of correlator is given by the expression conjectured in \cite{Bajnok:2014sza}. Apart from giving the general arguments of the proof we computed the simplest non-trivial correlators at tree level, which correspond to the case when the length of the light operator is equal to 4. The structure of the coefficients of the finite volume expansion allowed us to conjecture their form for any number of excitations of the heavy operator. We showed that these coefficients can be expressed in terms of several functions of the rapidities defined by the light operator. Another aspect of the conjecture proposed in \cite{Bajnok:2014sza} suggests that the coefficients of the above mentioned finite volume expansion can be interpreted as appropriate infinite volume form factors. Therefore our explicit calculation provides us with concrete proposal for the infinite volume form factor.

As a continuation of our work it would be interesting to see how the coefficients in the finite volume expansion are deformed at one loop. As an obstacle, one should mention that, as it follows from section 9, the complexity of the simplest one-loop computation is equivalent to a tree level computation with a light operator of the length equal to 6, and the calculation becomes much more involved. On the other hand, the recent result of \cite{Basso:3ptf} suggests an all-loop method for computing three-point correlators. And it would be interesting to merge their method with the approach developing in this paper and check the conjecture of \cite{Bajnok:2014sza} at all loop. In addition, it would give us all-loop prediction for the diagonal infinite volume form factors.

Another interesting direction to pursue is to try to see whether the same structure of the finite volume dependence holds for symmetric HHL correlators of the other sectors of ${\cal N}$=4 SYM.

\acknowledgments
It is our pleasure to thank R. Janik, I. Kostov and D. Serban for numerous helpful discussions. We are especially grateful to
Z. Bajnok for the various discussions and valuable comments on the manuscript.
Y.J. and A.P. would like to thank YITP at Stony Brook, Y.J. to the Wigner Research Centre of Physics at Budapest where a part of the work was done. L.H. is grateful for
the hospitality of the Institut de Physique Th\'eorique, Saclay and
the Laboratoire de Physique Th\'eorique, \'Ecole Normale Sup\'erieure.
Y.J. and A.P. has received support
from the European Programme IRSES UNIFY (Grant No 269217) and from the
People Programme (Marie Curie Actions) of the European Union's Seventh
Framework Programme FP7/2007-2013/ under REA Grant Agreement No 317089. L.H.
was supported by the Lend\"ulet Grant LP2012-18/2015, and the French-Hungarian bilateral
grant TET 12 FR-1-2013-0024.

\appendix

\section{An example of finite volume expansion}

\label{Appendix1} In this appendix, we give an explicit example in
order to illustrate how to perform the finite volume expansion of
the special scalar products defined in section \ref{sec:FVE}. The
scalar product under consideration is $\langle\{u_{1},u_{2},u_{3},u_{4}\}|\{u_{1},\theta,u_{3},u_{4}\}\rangle$,
where $\{u_{1},u_{2},u_{3},u_{4}\}$ is a set of Bethe roots. Consider
the numerator of the Slavnov determinant formula (\ref{sp}),
\begin{align}
\langle\{u_{1},u_{2},u_{3},u_{4}\}|\{u_{1},\theta,u_{3},u_{4}\}\rangle\propto\left|\begin{array}{cccc}
\phi_{11} & \Omega_{12} & \phi_{13} & \phi_{14}\\
\phi_{21} & \Omega_{22} & \phi_{23} & \phi_{24}\\
\phi_{31} & \Omega_{32} & \phi_{33} & \phi_{34}\\
\phi_{41} & \Omega_{42} & \phi_{43} & \phi_{44}
\end{array}\right|
\end{align}
We first perform the Laplace expansion for the second column, which
gives
\begin{align}
 & \,-\Omega_{12}\left|\begin{array}{ccc}
\phi_{21} & \phi_{23} & \phi_{24}\\
\phi_{31} & \phi_{33} & \phi_{34}\\
\phi_{41} & \phi_{43} & \phi_{44}
\end{array}\right|+\Omega_{22}\left|\begin{array}{ccc}
\phi_{11} & \phi_{13} & \phi_{14}\\
\phi_{31} & \phi_{33} & \phi_{34}\\
\phi_{41} & \phi_{43} & \phi_{44}
\end{array}\right|\\
 & \,-\Omega_{32}\left|\begin{array}{ccc}
\phi_{11} & \phi_{13} & \phi_{14}\\
\phi_{21} & \phi_{23} & \phi_{24}\\
\phi_{41} & \phi_{43} & \phi_{44}
\end{array}\right|+\Omega_{42}\left|\begin{array}{ccc}
\phi_{11} & \phi_{13} & \phi_{14}\\
\phi_{21} & \phi_{23} & \phi_{24}\\
\phi_{31} & \phi_{33} & \phi_{34}
\end{array}\right|\nonumber
\end{align}
The Laplace expansion gives rise to $4$ terms, which we shall denote
$\rT_{i}$, $i=1,\cdots,4$. For $\rT_{1}$, we do Laplace expansion
by the first column
\begin{align*}
\rT_{1}= & \,-\Omega_{12}\left\{ \phi_{21}\rho_{4}(\{3,4\})-\phi_{31}(\phi_{23}\rho_{4}(\{4\})-\phi_{24}\phi_{43})+\phi_{41}(\phi_{23}\phi_{34}-\phi_{24}\rho_{4}(\{3\}))\right\} \\\nonumber
= & \,-\Omega_{12}\phi_{21}\,\rho_{4}(\{3,4\})+\Omega_{12}\phi_{41}\phi_{23}\,\rho_{4}(\{3\})+\Omega_{12}\phi_{31}\phi_{23}\,\rho_{4}(\{4\})\\\nonumber
&\,-\Omega_{12}(\phi_{31}\phi_{24}\phi_{43}+\phi_{41}\phi_{24}\phi_{34})\nonumber
\end{align*}
The second term already takes the form of diagonal minor of the Gaudin
norm
\begin{align}
\rT_{2}=\Omega_{22}\,\,\rho_{4}(\{1,3,4\})
\end{align}
For the third term, we perform Laplace expansion with respect to the
second column
\begin{align*}
\rT_{3}=&\,-\Omega_{32}\phi_{23}\,\rho_{4}(\{1,4\})+\Omega_{32}\phi_{43}\phi_{24}\,\rho_{4}(\{1\})+\Omega_{32}\phi_{13}\phi_{21}\,\rho_{4}(\{4\})\\
&\,-\Omega_{32}(\phi_{13}\phi_{24}\phi_{41}+\phi_{43}\phi_{14}\phi_{21})
\end{align*}
For the last term, we perform Laplace expansion with respect to the
last column
\begin{align*}
\rT_{4}=&\,-\Omega_{42}\phi_{24}\,\rho_{4}(\{1,3\})+\Omega_{42}\phi_{34}\phi_{23}\,\rho_{4}(\{1\})+\Omega_{42}\phi_{14}\phi_{21}\,\rho_{4}(\{3\})\\
&\,-\Omega_{42}(\phi_{14}\phi_{23}\phi_{31}+\phi_{34}\phi_{13}\phi_{21})
\end{align*}
Collecting terms from the above calculation, we obtain the finite
volume expansion of the scalar product
\begin{align}
&\,\langle\{u_{1},u_{2},u_{3},u_{4}\}|\{u_{1},\theta,u_{3},u_{4}\}\rangle\propto \\\nonumber
 & \,\Omega_{22}\,\rho_{4}(\{1,3,4\})-\Omega_{42}\phi_{24}\,\rho_{4}(\{1,3\})-\Omega_{32}\phi_{23}\,\rho_{4}(\{1,4\})-\Omega_{12}\phi_{21}\,\rho_{4}(\{3,4\})\\
 & \,+(\Omega_{32}\phi_{43}\phi_{24}+\Omega_{42}\phi_{34}\phi_{23})\,\rho_{4}(\{1\})+(\Omega_{42}\phi_{14}\phi_{21}+\Omega_{12}\phi_{41}\phi_{24})\,\rho_{4}(\{3\})\nonumber \\
 & \,+(\Omega_{12}\phi_{31}\phi_{23}+\Omega_{32}\phi_{13}\phi_{21})\,\rho_{4}(\{4\})-\Omega_{12}(\phi_{31}\phi_{24}\phi_{43}+\phi_{41}\phi_{23}\phi_{34})\nonumber \\
 & \,-\Omega_{32}(\phi_{13}\phi_{24}\phi_{41}+\phi_{43}\phi_{14}\phi_{21})-\Omega_{42}(\phi_{14}\phi_{23}\phi_{31}+\phi_{34}\phi_{13}\phi_{21})\nonumber
\end{align}
In fact, it is not hard to convince ourselves that the similar expansion
can be performed for general scalar products defined in section \ref{sec:FVE}.
For length-$2$ operators, we have the following finite volume expansion
\begin{align}
\langle\mathbf{u}|\{\mathbf{u},\theta_{n},\theta_{n+1}\}\setminus\{u_{j},u_{k}\}\rangle=\sum_{\alpha\subseteq A}F_{\bar{\alpha}}\,\rho(\alpha),\quad A=\{1,2,...,\hat{j},...,\hat{k},...,N\}
\end{align}
In general, the terms of the expansion grows quickly with the number
of excitations and the expansion coefficients $F_{\bar{\alpha}}$
might get quite involved.

\providecommand{\href}[2]{#2}\begingroup\raggedright\endgroup

\end{document}